\documentclass[aps,prd,twocolumn,showpacs,floats,floatfix,letterpaper,
nofootinbib,superscriptaddress,preprintnumbers]{revtex4-1}
\usepackage{amssymb,amsmath,latexsym,mathrsfs}
\usepackage{graphicx}
\graphicspath{{images/}}
\usepackage{epsfig}
\usepackage{multirow}
\usepackage{array}
\usepackage{xspace}

\usepackage{hyperref}

\newcommand{\camb}{\texttt{CAMB}\xspace}
\newcommand{\cosmomc}{\texttt{CosmoMC}\xspace}
\newcommand{\lcdm}{\ensuremath{\Lambda\textrm{CDM}}\xspace}
\newcommand{\esp}[1]{\ensuremath{\times 10^{#1}}}
\newcommand{\eff}{\ensuremath{_{\rm{eff}}}}
\newcommand{\neff}{\ensuremath{N_{\rm{eff}}}}
\newcommand{\meff}{\ensuremath{m_{s}^{\rm{eff}}}}

\newcommand{\mpcinv}{\ensuremath{\, \text{Mpc}^{-1}}}
\newcommand{\psj}[1]{\ensuremath{P_{s,#1}}}
\newcommand{\mnu}{\ensuremath{\Sigma m_\nu}}

\newcommand{\pchip}{\texttt{PCHIP}\xspace}

\renewcommand{\thefootnote}{\arabic{footnote}}

\begin{document}

\def\thefootnote{\fnsymbol{footnote}}

\title{Dark Radiation and Inflationary Freedom after Planck 2015}

\author{Eleonora Di Valentino}
\affiliation{Institut d'Astrophysique de Paris (UMR7095: CNRS \& UPMC- Sorbonne Universities), F-75014, Paris, France}

\author{Stefano Gariazzo}
\affiliation{Department of Physics, University of Torino,
Via P. Giuria 1, I--10125 Torino, Italy}
\affiliation{INFN, Sezione di Torino,
Via P. Giuria 1, I--10125 Torino, Italy}

\author{Martina Gerbino} 
\affiliation{The Oskar Klein Centre for Cosmoparticle Physics, Department of Physics, Stockholm University, AlbaNova, SE-106 91 Stockholm, Sweden}
\affiliation{Nordita (Nordic Institute for Theoretical Physics), Roslagstullsbacken 23, SE-106 91 Stockholm, Sweden}
\affiliation{Physics Department and INFN, Universit\`a di Roma 
	``La Sapienza'', P.le\ Aldo Moro 2, 00185, Rome, Italy}

\author{Elena Giusarma} 
\affiliation{Physics Department and INFN, Universit\`a di Roma 
	``La Sapienza'', P.le\ Aldo Moro 2, 00185, Rome, Italy}
\affiliation{McWilliams Center for Cosmology, Department of Physics,
Carnegie Mellon University, Pittsburgh, PA 15213, USA}

\author{Olga Mena} 
\affiliation{IFIC, Universidad de Valencia-CSIC, 46071, Valencia, Spain}

\begin{abstract}

{The simplest inflationary models predict
a primordial power spectrum (PPS) of the curvature fluctuations
that can be described by a power-law function that is nearly scale-invariant.
It has been shown, however, that the low-multipole spectrum
of the CMB anisotropies may hint the presence of some features in the shape of the scalar PPS, 
which could deviate from its canonical power-law form.
We study the possible degeneracies of this non-standard PPS with the
active neutrino masses,
the effective number of relativistic species and a sterile neutrino or a thermal axion mass.
The limits on these additional parameters are less constraining in a model with a non-standard PPS
when including only the temperature auto-correlation spectrum measurements in the data analyses. The inclusion of the polarization spectra noticeably helps in reducing the degeneracies, leading to results that typically show no deviation
from the \lcdm\ model with a standard power-law PPS. These findings are robust against changes in the function describing the non-canonical PPS. Albeit current cosmological measurements seem to prefer the simple power-law PPS description, the statistical significance to rule out other possible parameterizations is still very poor. Future cosmological measurements are crucial to improve the present PPS uncertainties.}

\end{abstract}


\maketitle

\section{Introduction}
\label{sec:intro}
Inflation is one of the most successful theories
that explains the so-called \emph{horizon} and \emph{flatness}
problems, providing an origin for the primordial density perturbations
that evolved to form the structures we observe today~\cite{Guth:1980zm,Linde:1981mu,Starobinsky:1982ee,Hawking:1982cz,
Albrecht:1982wi,Mukhanov:1990me,Mukhanov:1981xt,Lucchin:1984yf,
Lyth:1998xn,Bassett:2005xm,Baumann:2008bn}.
The standard inflationary paradigm predicts a simple shape
for the primordial power spectrum (PPS) of scalar perturbations:
in this context, the PPS can be described by a power-law expression.
However, there also exist more complicated inflationary scenarios which can give rise to non-standard PPS forms,
with possible features at different scales, see e.g.~Refs.~\cite{Romano:2014kla,Kitazawa:2014dya}
and the reviews \cite{Martin:2013tda,Chluba:2015bqa}.

The usual procedure to reconstruct the underlying PPS is to assume a model
for the evolution of the Universe and calculate the transfer function, 
 and then use different techniques to constrain
a completely unknown PPS,
comparing the theoretical prediction with 
the measured power spectrum of the
Cosmic Microwave Background radiation (CMB).
Among the methods developed in the past, we can list
regularization methods as the Richardson-Lucy iteration
\cite{Shafieloo:2003gf,Nicholson:2009pi,Hazra:2013eva,Hazra:2014jwa},
truncated singular value decomposition \cite{Nicholson:2009zj}
and Tikhonov regularization \cite{Hunt:2013bha,Hunt:2015iua},
or methods as the
maximum entropy deconvolution \cite{Goswami:2013uja}
or the \emph{cosmic inversion} methods
\cite{Matsumiya:2001xj,Matsumiya:2002tx,Kogo:2003yb,
Kogo:2005qi,Nagata:2008tk}. Recently, the Planck collaboration presented a wide discussion 
about constraints on
inflation~\cite{Ade:2015lrj}.
All these methods provide hint for a PPS which may not be
as simple as a power-law. While the significance of the deviations is small for some cases, it is interesting to note that the CMB temperature power spectra as measured by both WMAP \cite{Bennett:2012zja}
and Planck \cite{Ade:2013sjv, Adam:2015rua} show similar results: the differences from the power-law
 are located in the low multipole region. These deviations could arise from some statistical fluctuations, or, instead, result from a non-standard inflationary mechanism.
\newline 
If the features we observe are the result of
a non-standard inflationary mechanism,
we may be using an incomplete parameterization
for the PPS in our cosmological analyses.
It has been shown that this could lead to biased results 
in the cosmological constraints of different quantities. Namely, the constraints on the dark radiation properties
\cite{dePutter:2014hza,Gariazzo:2014dla, DiValentino:2015zta}
or on non-Gaussianities \cite{Gariazzo:2015qea} can be distorted, 
leading to spurious conclusions. In this work we aim to study the
impact of a general PPS form in the constraints obtained for the
properties of dark radiation candidates, such as the active neutrino
masses and their effective number, sterile neutrino species and thermal axion properties.
The outline of the Paper is as follows:
we present the baseline standard \lcdm~ cosmological model, the PPS parameterization
and the cosmological data in Sec.~\ref{sec:common}. The results obtained within the \lcdm~framework are presented in Sec.~\ref{sec:lcdm}. Concerning possible extensions of the \lcdm\ scenario,
we study the constraints on the effective number of relativistic species in Sec.~\ref{sec:nnu},
on the neutrino masses in Sec.~\ref{sec:nu}, on massive neutrinos with
a varying effective number of relativistic species in
Sec.~\ref{sec:mnu_nnu}, on massive sterile neutrinos in
Sec.~\ref{sec:meff}, and on the thermal axion properties in Sec.~\ref{sec:ax}.
Finally, in Sec.~\ref{sec:pps} we show the reconstructed PPS shape, comparing different possible approaches, and we draw our conclusions in Sec.~\ref{sec:conclusions}.


\section{Baseline Model and Cosmological Data}\label{sec:common}
In this Section we outline the baseline theoretical model that will be
extended to study the dark radiation properties. 
For our analyses we use the numerical Boltzmann solver
\camb~\cite{Lewis:1999bs} for the theoretical spectra calculation, and the Markov Chain Monte Carlo (MCMC) algorithm
\cosmomc~\cite{Lewis:2002ah} to sample the parameter space.

\subsection{Standard Cosmological Model}\label{ssec:lcdm}
The baseline model that we will extend
to study various dark radiation properties
is the \lcdm~model, described by the six usual parameters:
the current energy density of baryons and of Cold Dark Matter (CDM)
($\Omega_b h^2$, $\Omega_c h^2$),
the ratio between the sound horizon and
the angular diameter distance at decoupling ($\theta$),
the optical depth to reionization ($\tau$),
plus two parameters that describe the PPS of scalar perturbations,
$P_s(k)$.
The simplest models of inflation predict a power-law form
for the PPS:
\begin{equation}\label{eq:pps_pl}
  P_s(k)=A_s \left(k/k_*\right)^{n_s-1}\,,
\end{equation}
where $k_*=0.05\mpcinv$ is the pivot scale,
while the amplitude ($A_s$) and
the scalar spectral index ($n_s$) are free parameters
in the \lcdm~model. From these fundamental cosmological parameters we will compute
other derived quantities, such as the Hubble parameter today $H_0$
and the clustering parameter $\sigma_8$,
defined as the mean of matter fluctuations inside a sphere of 8$h$~Mpc radius.

From what concerns the remaining cosmological parameters, we follow the values of Ref.~\cite{Ade:2015xua}.
In particular, unless they are freely varying, we consider the sum of
the active neutrino masses to be $\mnu=0.06$~eV, and the effective number of relativistic species to be
$\neff=3.046$ \cite{Mangano:2005cc}. 

\subsection{Primordial Power Spectrum of Scalar Perturbations}
\label{ssec:pps}
As stated before, possible hints
of a non-standard PPS of scalar perturbations
were found in several analyses, including both the WMAP and the Planck CMB spectra 
\cite{Shafieloo:2003gf,Nicholson:2009pi,Hazra:2013eva,Hazra:2014jwa,
Nicholson:2009zj,Hunt:2013bha,Hunt:2015iua,
Goswami:2013uja,Matsumiya:2001xj,Matsumiya:2002tx,
Kogo:2003yb,Kogo:2005qi,Nagata:2008tk,Ade:2015lrj,
Gariazzo:2014dla,DiValentino:2015zta}.
From the theoretical point of view,
there are plenty of well-motivated inflationary models
that can give rise to non-standard PPS forms. Our major goal here is
to study the robustness of the constraints on different cosmological
quantities versus a change in the assumed PPS.
Several cosmological parameters are known to
present degeneracies with the standard PPS parameters, 
as, for example, the existing one between effective number of relativistic species $\neff$
and the tilt of the power-law PPS $n_s$.  These degeneracies could be even stronger
when more freedom is allowed for the PPS shape. 
We adopt here a non-parametric description
for the PPS of scalar perturbations:
we describe the function $P_s(k)$
as the interpolation among a series of nodes at fixed wavemodes $k$.
Unless otherwise stated, we shall consider twelve nodes $k_j$ ($j\in[1,12]$)
that cover a wide range of values of $k$:
the most interesting range is explored between
$k_2=0.001\mpcinv$ and $k_{11}=0.35\mpcinv$, 
that is approximately the range of wavemodes probed
by CMB experiments. 
In this range we use equally spaced nodes in $\log k$.
Additionally, we consider $k_1=5\esp{-6}\mpcinv$ and 
$k_{12}=10\mpcinv$ 
in order to ensure that all the PPS evaluations are
inside the covered range. We expect that the nodes at these extreme wavemodes
are  unconstrained by the data.

Having fixed the position of all the nodes,
the free parameters that are involved in our MCMC analyses
are the values of the PPS at each node, $\psj{j}=P_s(k_j)/P_0$, 
where $P_0$ is the overall normalization,
$P_0=2.2\esp{-9}$~\cite{Ade:2015xua}.
We use a flat prior in the interval $[0.01,10]$ for each $\psj{j}$,
for which the expected value will be close to 1.

The complete $P_s(k)$ is then described as
the interpolation among the points $\psj{j}$:
\begin{equation}\label{eq:PPS_pchip}
  P_{s}(k)=P_0\times\pchip(k; \psj{1}, \ldots, \psj{12})\,,
\end{equation}
where \pchip\ is the
\emph{piecewise cubic Hermite interpolating
polynomial}~\cite{Fritsch:1980,Fritsch:1984} (see also
Ref.~\cite{Gariazzo:2014dla} for a detailed description\footnote{The
  PCHIP method is similar to the natural cubic spline,
but it has the advantage of avoiding the introduction
of spurious oscillations in the interpolation:
this is obtained with a condition
on the first derivative in the nodes,
that is null if there is a change in the monotonicity
of the point series.}).
In the following, when presenting our results,
we will compare the constraints obtained in the context of the 
standard \lcdm~model with a standard power-law  PPS
to those obtained with the free \pchip~PPS, described by (at least) sixteen free parameters
($\Omega_b h^2$, $\Omega_c h^2$, $\theta$, $\tau$,
$\psj{1},\ldots, \psj{12}$). This minimal model will be extended to include
the dark radiation properties we shall study in the various analyses.

The impact of the assumptions on the PPS parameterization will also be
tested. We shall compare the results obtained with twelve nodes to the ones derived
using a \pchip\ PPS described by eight nodes.
The position of these eight nodes $k_j^{(8)}$
is selected with the same rules as above:
equally spaced nodes in $\log k$ between
$k_2^{(8)}=k_2=0.001\mpcinv$ and $k_{7}^{(8)}=k_{11}=0.35\mpcinv$, plus
the external nodes $k_1^{(8)}=k_1=5\esp{-6}\mpcinv$ and 
$k_{8}^{(8)}=k_{12}=10\mpcinv$.

To ease comparison bewteen the power-law and the \pchip PPS approaches,
we list in all the tables the results obtained for these two schemes.
When considering a power-law PPS model, we show the constraints on $n_s$ and $A_s$,
together with the values of the nodes $\psj{1}^{\textrm{bf}}$
to $\psj{12}^{\textrm{bf}}$ that would correspond to the best-fit
values of $n_s$ and $A_s$ ($n_s^{\textrm{bf}}$ and
$A_s^{\textrm{bf}}$). In other words, in each table presenting the marginalized
constraints for the different cosmological parameters, in the columns corresponding to the analysis
involving a power-law PPS, we shall list the values
\begin{equation}\label{eq:psjbf}
  \psj{j}^{\textrm{bf}}
  \equiv
  \frac{A_s^{\textrm{bf}}}{P_0}
  \,
  \left(\frac{k_j}{k_*}\right)^{n_s^{\textrm{bf}}-1}
  \quad\mbox{ with }
  j\in [1,\ldots,12]
  \,, 
\end{equation}
that can be exploited for comparison purposes among the two PPS approaches.

\subsection{Cosmological data}\label{ssec:data}
We base our analyses on the recent release
from the Planck Collaboration \cite{Adam:2015rua},
that obtained the most precise
CMB determinations in a very wide range of multipoles.
We consider the full temperature power spectrum
at multipoles $2\leq\ell\leq2500$ (``Planck TT'' hereafter)
and the polarization power spectra in the range $2\leq\ell\leq29$ (``lowP'').
We shall also include the polarization data
at $30\leq\ell\leq2500$ (``TE, EE'') \cite{Aghanim:2015xee}.
Since the polarization spectra at high multipoles are still under discussion and some residual systematics
were detected by the Planck Collaboration
\cite{Aghanim:2015xee,Ade:2015xua},
we shall use as baseline dataset the combination
``Planck~TT+lowP'' . The impact of polarization measurements will be
separately studied in the dataset ``Planck~TT,TE,EE+lowP''.

Additionally, we will consider the two CMB datasets above in
combination with the following cosmological measurements:
\begin{itemize}
  \item \textbf{BAO}: Baryon Acoustic Oscillations data as obtained by
    6dFGS \cite{Beutler:2011hx} at redshift $z=0.1$, 
    by the SDSS Main Galaxy Sample (MGS) \cite{Ross:2014qpa}
    at redshift $z\eff=0.15$ and
    by the BOSS experiment in the DR11 release,
    both from the LOWZ and CMASS samples \cite{Anderson:2013zyy} 
    at redshift $z\eff=0.32$ and $z\eff=0.57$, respectively;
  \item \textbf{MPkW}: the matter power spectrum as measured by the 
    WiggleZ Dark Energy Survey \cite{Parkinson:2012vd}, 
    from measurements at four different redshifts
    ($z=0.22$, $z=0.41$, $z=0.60$ and $z=0.78$) for the scales $0.02\,h\,\mathrm{Mpc^{-1}}<k<0.2\,h\,\mathrm{Mpc^{-1}}$;
  \item \textbf{lensing}: the reconstruction of the lensing potential
    obtained by the Planck collaboration
    with the CMB trispectrum analysis \cite{Ade:2015zua}.
\end{itemize}

\section{The \texorpdfstring{\lcdm}{LambdaCDM} Model}\label{sec:lcdm}
\begin{figure}
\centering
\includegraphics[width=0.9\columnwidth,page=1]{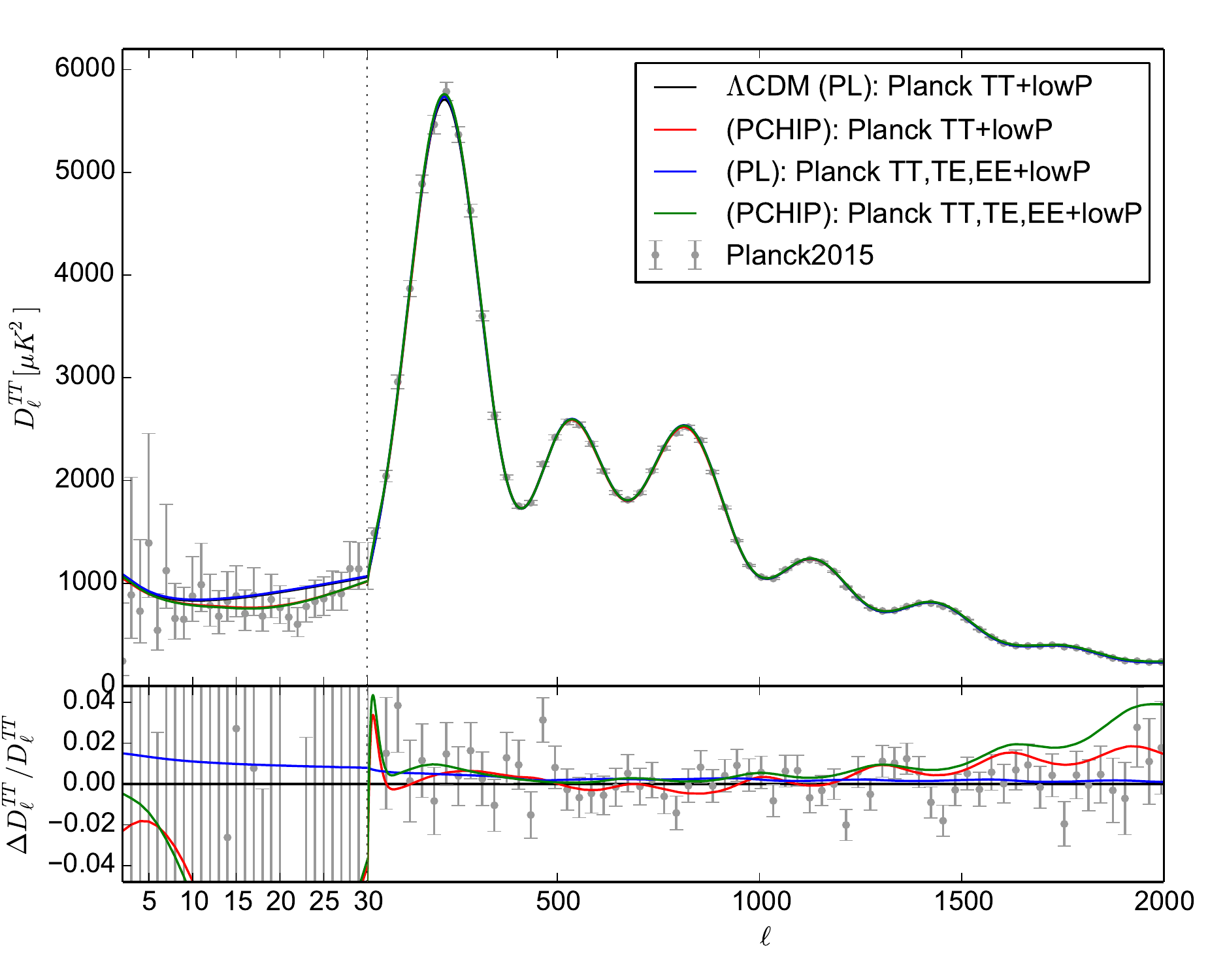}
\includegraphics[width=0.9\columnwidth,page=2]{images/cl_lcdm.pdf}
\includegraphics[width=0.9\columnwidth,page=3]{images/cl_lcdm.pdf}
\caption{
 Comparison of the Planck 2015 data \cite{Adam:2015rua}
 with the TT, TE and EE spectra obtained using the marginalized
 best-fit values from the analyses of 
 Planck~TT+lowP (black) and 
 Planck~TT,TE,EE+lowP (blue) in the \lcdm\ model with the power-law (PL) PPS,
 and from the analyses of 
 Planck~TT+lowP (red) and 
 Planck~TT,TE,EE+lowP (green) in the \lcdm\ model with the \pchip\ PPS.
 The adopted values for each spectrum are reported in Tab.~\ref{tab:lcdm}.
 We plot the $D_\ell=\ell(\ell+1)\,C_\ell/(2\pi)$ spectra
 and the relative (absolute for the case of the TE spectra) difference
 between each spectrum and the one obtained
 in the \lcdm\ (power-law PPS) model from the Planck~TT+lowP data (black line).}
\label{fig:cl_lcdm}
\end{figure}

In this section we shall consider a limited number of data combinations, including exclusively the datasets that can improve the constraints on the \pchip PPS at small scales, namely, the Planck polarization measurements at high-$\ell$ and
the MPkW constraints on the matter power spectrum.

The results we obtain for the \lcdm\ model are reported in
Tab.~\ref{tab:lcdm} in the Appendix. In general, in the absence of high multipole polarization or large scale structure data, the parameter errors are increased.
Those associated to $\Omega_b h^2$, $\Omega_c h^2$, $H_0$ and $\sigma_8$ show a larger difference, with deviations of the order of 1$\sigma$
in the \pchip PPS case with respect to the power-law PPS case.  The differences between the \pchip and the power-law PPS parameterizations are much smaller for the ``Planck TT,TE,EE+lowP+MPkW'' dataset, and the two descriptions of the PPS
give bounds for the \lcdm\ parameters tha fully agree. Therefore, the addition of the high multipole polarization spectra has a profound impact in our analyses, as we carefully explain in what follows. 
Figure~\ref{fig:cl_lcdm} depicts the CMB spectra measured by Planck
\cite{Adam:2015rua}, together with the theoretical spectra obtained from the best-fit
values arising from our analyses. More concretely, we use the marginalized best-fit values reported in Tab.~\ref{tab:lcdm}
for the \lcdm\ model with a power-law PPS obtained from the analyses of the
Planck~TT+lowP (in black) and Planck~TT,TE,EE+lowP (in blue) datasets,
plus the best-fit values in the \lcdm\ model with a \pchip\ PPS,
from the Planck~TT+lowP (red) and Planck~TT,TE,EE+lowP (green) datasets.
We plot the $D_\ell=\ell(\ell+1)\,C_\ell/(2\pi)$ spectra
of the TT, TE and EE anisotropies, as well as the relative (absolute for the TE spectra) difference
between each spectrum and the one obtained
from the Planck~TT+lowP data in the \lcdm\ model with the power-law PPS.
Notice that, in the case of the  TT and EE spectra, the best-fit
spectra are in good agreement with the observational data, even if
there are variations among the \lcdm\ parameters, 
 as they can be compensated by the freedom in the PPS.
However, in the TE cross-correlation spectrum case, such a
compensation is no longer possible: the inclusion of the TE spectrum
in the analyses is therefore expected to have a strong impact on the derived bounds.

\section{Effective Number of Relativistic Species}\label{sec:nnu}
The amount of energy density of relativistic species
in the Universe 
is usually defined as the sum of the photon contribution
$\rho_\gamma$
plus the contribution of all the other relativistic species.
This is described by the effective number of relativistic degrees of freedom \neff:
\begin{equation}\label{eq:rhorad}
 \rho_{\mathrm{rad}} = \left[1 + \frac{7}{8}
    \left(\frac{4}{11}\right)^{4/3}\neff\right]\rho_{\gamma} \, ,
\end{equation}
where $\neff=3.046$ \cite{Mangano:2005cc} for the three active neutrino standard scenario. 
Deviations of $\neff$ from its standard value may indicate that the thermal history of the active neutrinos
is different from what we expect, or that additional relativistic particles are present in the Universe, as 
additional sterile neutrinos or thermal axions.

A non-standard value of $\neff$ may affect the Big Bang Nucleosynthesis era, and also the matter-radiation equality. A shift in the matter-radiation equality would cause a change in the expansion rate at decoupling, affecting the sound horizon and the angular scale of the peaks of the CMB spectrum, as well as in the contribution of the \emph{early Integrated Sachs
Wolfe (ISW) effect} to the CMB spectrum. To avoid such a shift and its
consequences, it is possible to change simultaneously the energy
densities of matter and dark energy, in order to keep fixed all the relevant scales
in the Universe. In this case, the CMB spectrum will only be altered by an increased Silk damping at small scales
(see e.g.\ Refs.~\cite{Hou:2011ec,Lesgourgues-Mangano-Miele-Pastor-2013,Archidiacono:2013fha,Gariazzo:2015rra}).

\begin{figure}
\centering
\includegraphics[width=\columnwidth]{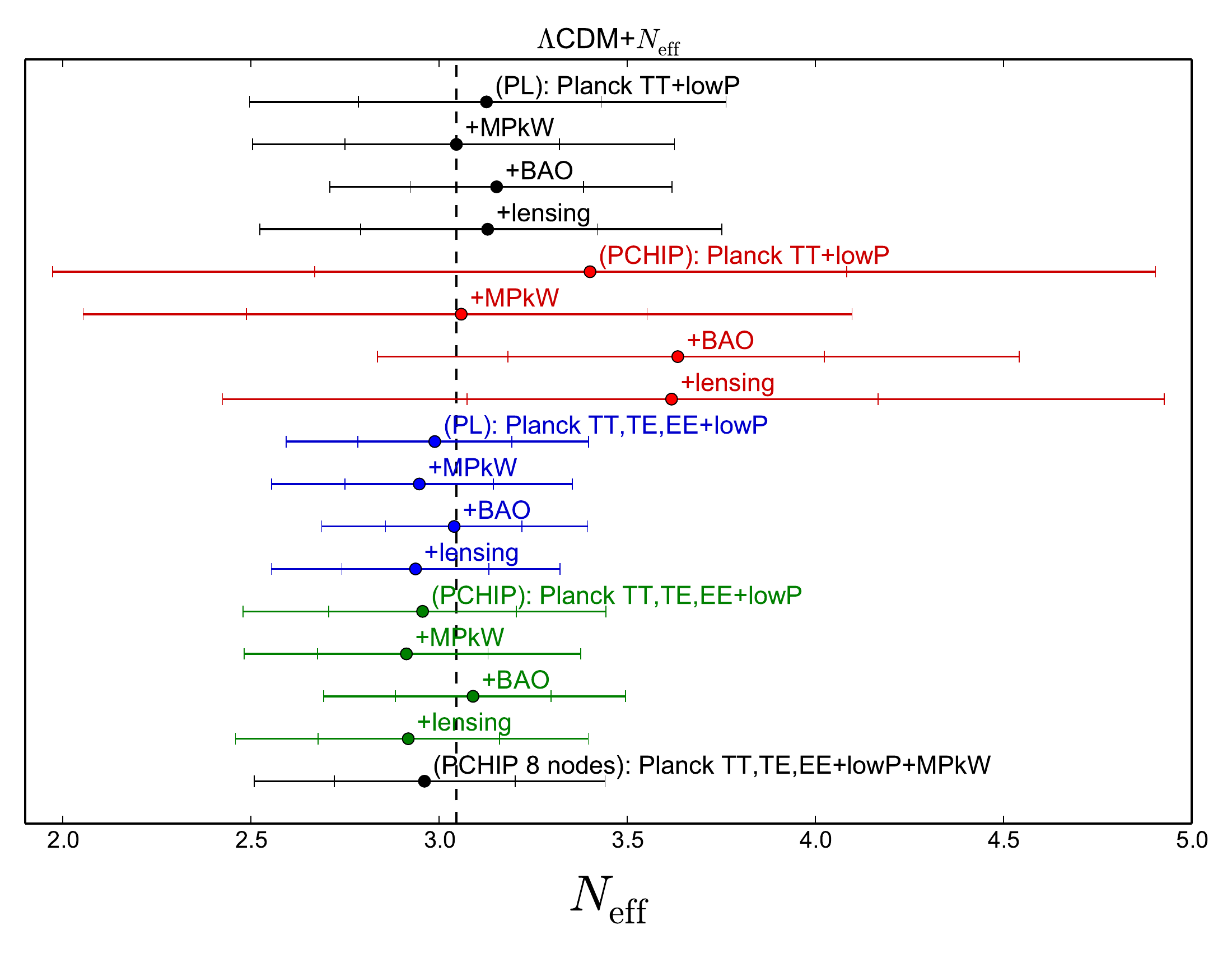}
\caption{68\% and 95\% CL constraints on $\neff$, obtained in the
 \lcdm + \neff\ model.
 Different colors indicate
 Planck~TT+lowP with PL PPS (black),
 Planck~TT+lowP with PCHIP PPS (red),
 Planck~TT,TE,EE+lowP with PL PPS (blue) and
 Planck~TT,TE,EE+lowP with PCHIP PPS (green).
 For each color we plot 4 different datasets: from top to bottom, we have
 CMB only, CMB+MPkW, CMB+BAO and CMB+lensing.
  We also illustrate the results, in the context of the 8-nodes parameterization,
  for the Planck~TT,TE,EE+lowP+MPkW dataset (last point in black).}
\label{fig:nnu_bars}
\end{figure}

The constraints on $\neff$ are summarized in Fig.~\ref{fig:nnu_bars},
where we plot the 68\% and 95\% CL constraints on $\neff$ obtained
with different datasets and PPS combinations for the \lcdm + \neff\ model.

\begin{figure*}
\centering
\includegraphics[width=\textwidth,page=1]{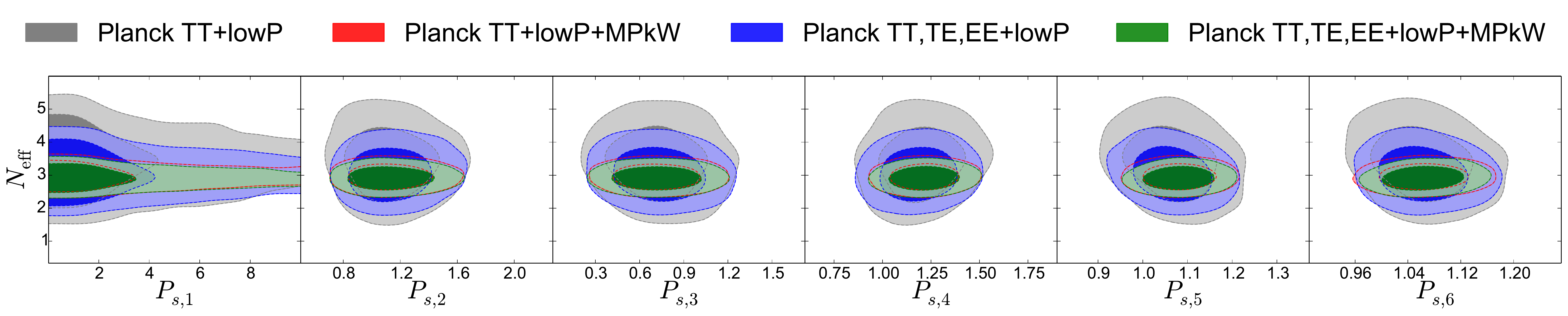}
\includegraphics[width=\textwidth,page=2]{images/nnu_corr.pdf}
\caption{68\% and 95\% CL constraints in the ($\neff$, \psj{j}) planes, obtained in the \lcdm + \neff\ model.
 We show the results for 
 Planck~TT+lowP (gray),
 Planck~TT+lowP+MPkW (red),
 Planck~TT,TE,EE+lowP (blue) and
 Planck~TT,TE,EE+lowP+MPkW (green).}
\label{fig:nnu_corr}
\end{figure*}

The introduction of $\neff$ as a free parameter does not change
significantly the results for the \lcdm\ parameters if a power-law PPS
is considered. However, once the freedom in the PPS is introduced,
some degeneracies between the \pchip\  nodes \psj{j} and \neff\
appear. Nevertheless, even if the constraints on \neff\ are loosened for the \pchip\ PPS case, all the dataset combinations give constraints on \neff\ that are compatible with the standard value 3.046 at 95\% CL, 
as we notice from  Fig.~\ref{fig:nnu_bars} and Tab.~\ref{tab:nnu}
in the Appendix. The mild preference for $\neff>3.046$ arises mainly
as a volume effect in the Bayesian analysis, since the \pchip\ PPS
parameters can be  tuned to reproduce the observed CMB temperature
spectrum for a wide range of values of \neff. As expected, the degeneracy between the nodes \psj{j} and \neff\ shows up at high wavemodes, where the Silk damping effect is dominant, see Fig.~\ref{fig:nnu_corr}.  As a consequence of this correlation, the values preferred for the nodes \psj{6} to \psj{10} are slightly larger than the  best-fit values in the power-law PPS at the same wavemodes. The cosmological limits for a number of parameters change as a consequence of the various degeneracies with \neff. For example, to compensate the shift of the matter-radiation equality redshift due to the increased radiation energy density, the CDM energy density $\Omega_c h^2$ mean value is slightly shifted and its constraints are weakened. At the same time, the uncertainty on the Hubble parameter $H_0$ is considerably relaxed, because $H_0$ must be also changed accordingly.  The introduction of the polarization data helps in improving the constraints in the models with a \pchip\ PPS,
since the effects of increasing \neff\ and changing the PPS are
different for the temperature-temperature,
the temperature-polarization and the polarization-polarization
correlation spectra, as previously discussed in the context of the
\lcdm\ model (see Tab.~\ref{tab:nnu_pol} in the Appendix): the preferred value of \neff\ is
very close to the standard value 3.046. Apparently, the Planck polarization data seem to  prefer 
a value of \neff\ slightly smaller than 3.046 for all the datasets except those including the BAO data, but the effect is not
statistically significant (see the blue and green points in Fig.~\ref{fig:nnu_bars}).

In conclusion, as the bounds for \neff\ are compatible with 3.046, the \lcdm + \neff\ model gives results that are very close to those obtained in the simple \lcdm\ model, but with slightly larger parameter uncertainties, in particular for $H_0$ and $\Omega_c h^2$.

\section{Massive Neutrinos}\label{sec:nu}

\begin{figure}
\centering
\includegraphics[width=\columnwidth,page=1]{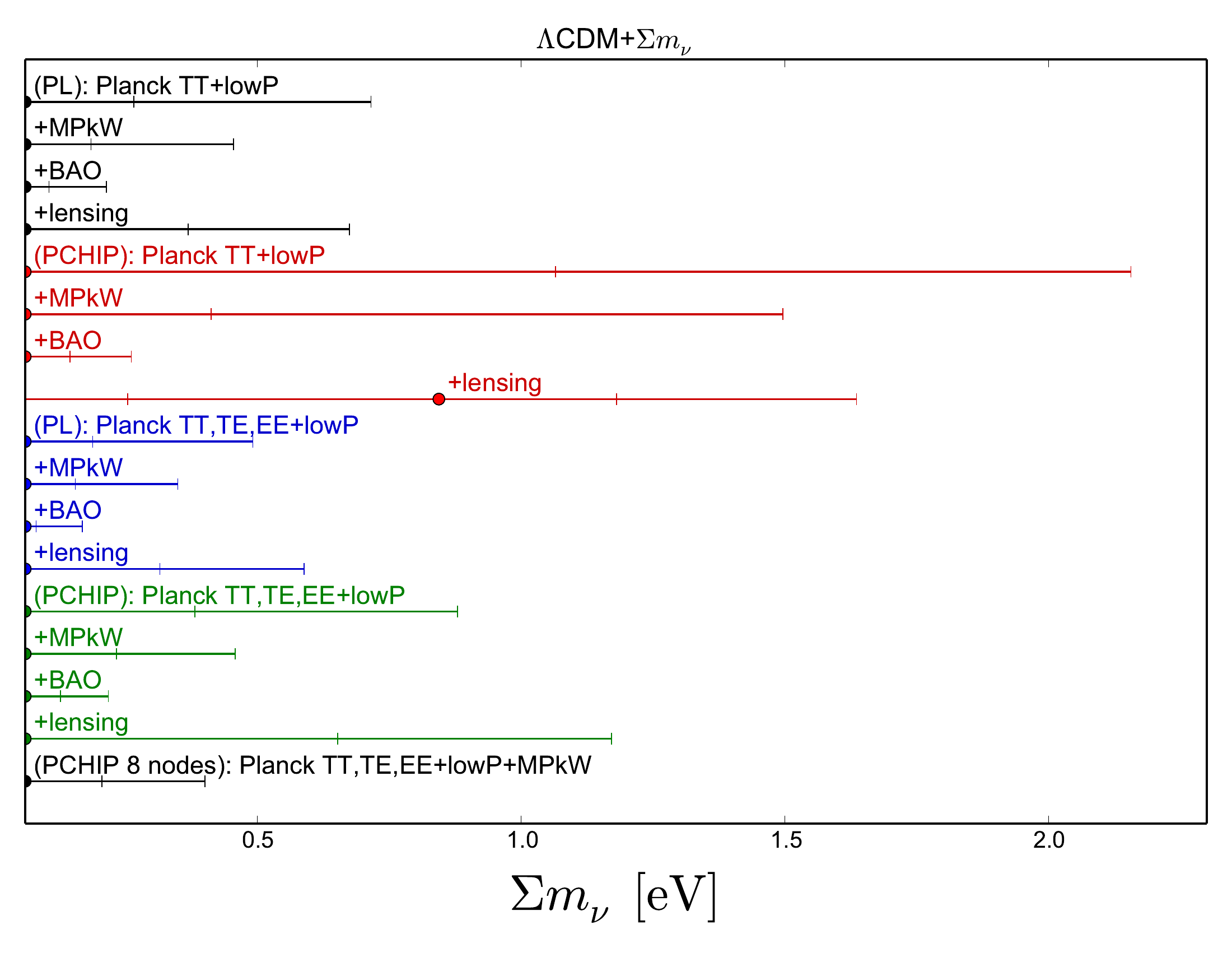}
\caption{As Fig.~\ref{fig:nnu_bars} but for the \lcdm plus $\sum m_\nu$ case.}
\label{fig:mnu_bars}
\end{figure}

Neutrinos oscillations have robustly established the existence of neutrino masses. However, neutrino mixing data only provide information on the squared mass differences and not on the absolute scale of neutrino masses. Cosmology provides an independent tool to test it, as massive neutrinos leave a non negligible imprint in different cosmological observables \cite{Reid:2009nq,Hamann:2010pw,dePutter:2012sh,Giusarma:2012ph,Zhao:2012xw,Hou:2012xq,Archidiacono:2013lva,Giusarma:2013pmn,Riemer-Sorensen:2013jsa,Hu:2014qma,Giusarma:2014zza,DiValentino:2015sam}.  The primary effect of neutrino masses in the CMB temperature spectrum is due to the early ISW effect. The neutrino transition from the relativistic to the non-relativistic regime affects the decay of the gravitational potentials at the decoupling period, producing an enhancement of the small-scale perturbations, especially near the first acoustic peak. A non-zero value of the neutrino mass also induces a higher expansion rate, which suppresses the lensing potential and the clustering on scales smaller than the horizon when neutrinos become non-relativistic.  
However, the largest effect of neutrino masses on the different cosmological observables comes from the suppression
of galaxy clustering at small scales. After becoming non-relativistic, the neutrino hot dark matter relics possess large velocity dispersions, suppressing the growth of matter density fluctuations at small scales. The baseline scenario we analyze here has three active massive neutrino species with degenerate masses. In addition, we consider the PPS approach outlined in Sec.~\ref{sec:common}.  For the numerical analyses,  when considering the power-law PPS, we use the following set of parameters:
\begin{equation}\label{eq:mnu_parameter}
\{\Omega_b h^2,\Omega_c h^2, \theta, \tau, n_s, \log[10^{10}A_{s}], \mnu\} ~.
\end{equation}

We then replace the $n_s$ and $A_s$ parameters with the other twelve extra parameters ($\psj{i}$ with $i=1,\ldots,12$) related to the \pchip PPS parameterization. The 68\% and 95\% CL bounds on \mnu\ obtained with different dataset and PPS combinations are summarized in Fig.~\ref{fig:mnu_bars}. 

\begin{figure*}
\centering
\includegraphics[width=\textwidth]{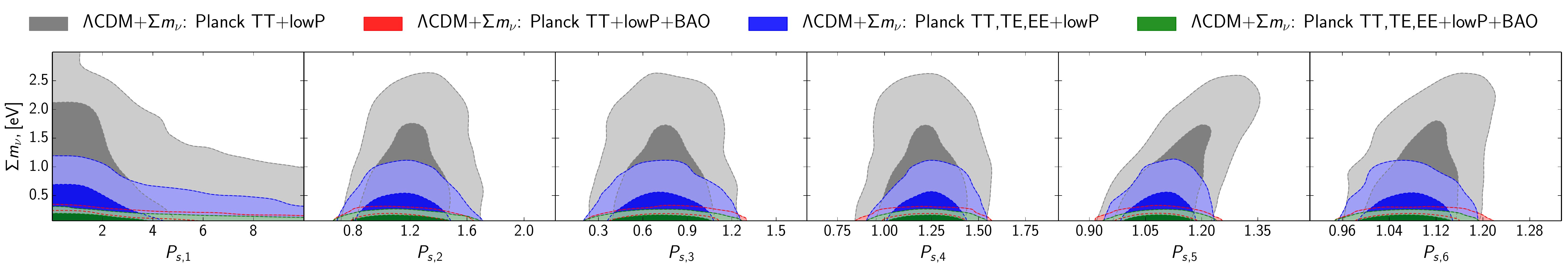}
\includegraphics[width=\textwidth]{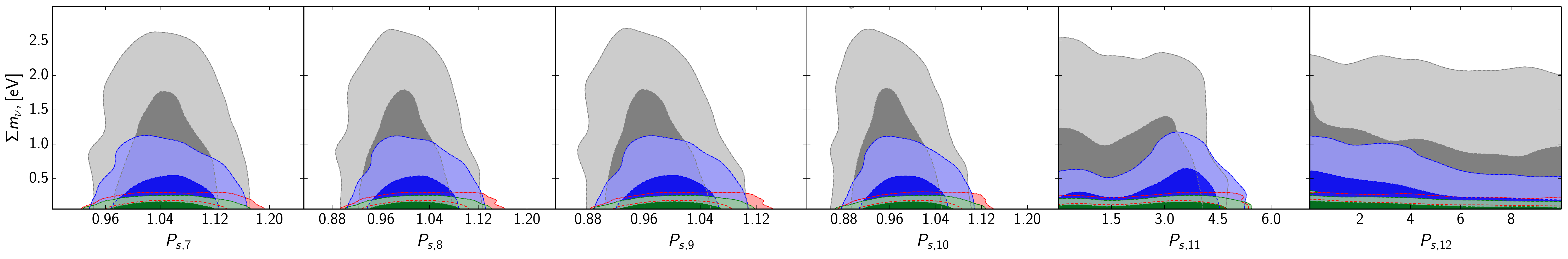}
\caption{As Fig.~\ref{fig:nnu_corr} but for the \lcdm plus $\sum m_\nu$ case.}
\label{fig:mnu_corr}
\end{figure*}

Notice that, when considering Planck TT+lowP CMB measurements plus other external datasets,  for all the data combinations, the bounds on neutrino masses are weaker when considering the \pchip PPS with respect to the power-law PPS case (see also Tab.~\ref{tab:mnuTT} in Appendix). Concerning CMB data only, the bound we find in the~\pchip approach is $\mnu<2.16$~eV at 95\% CL, much less constraining than the bound $\mnu<0.75$~eV at 95\% CL obtained in the power-law approach. This larger value is due to the degeneracy between \mnu\ and the nodes $\psj{5}$ and $\psj{6}$, as illustrated in Fig.~\ref{fig:mnu_corr}. In particular, these two nodes correspond to the wavenumbers where the contribution of the early ISW effect is located. Therefore, the change induced on these angular scales by a larger neutrino mass could be compensated by increasing $\psj{5}$ and $\psj{6}$. The addition of the matter power spectrum measurements, MPkW, leads to an upper bound on  \mnu\ of $1.15$~eV at 95\% CL in the~\pchip parameterization,  which is twice the value obtained when considering the power-law PPS with the same dataset.
The most stringent constraints on the sum of the three active neutrino masses are obtained when we use the BAO data, 
since the geometrical information they provide helps breaking degeneracies among cosmological parameters. In particular, we have $\mnu<0.261$~eV ($\mnu<0.220$~eV) at 95\% CL when considering the~\pchip (power-law) PPS parameterization. Finally, the combination of Planck TT+lowP data with the Planck CMB lensing measurements provide a bound on neutrino masses of $\mnu<1.64$~eV at 95\% CL in the~\pchip case.

\begin{figure}[t]
\includegraphics[width=8.5cm]{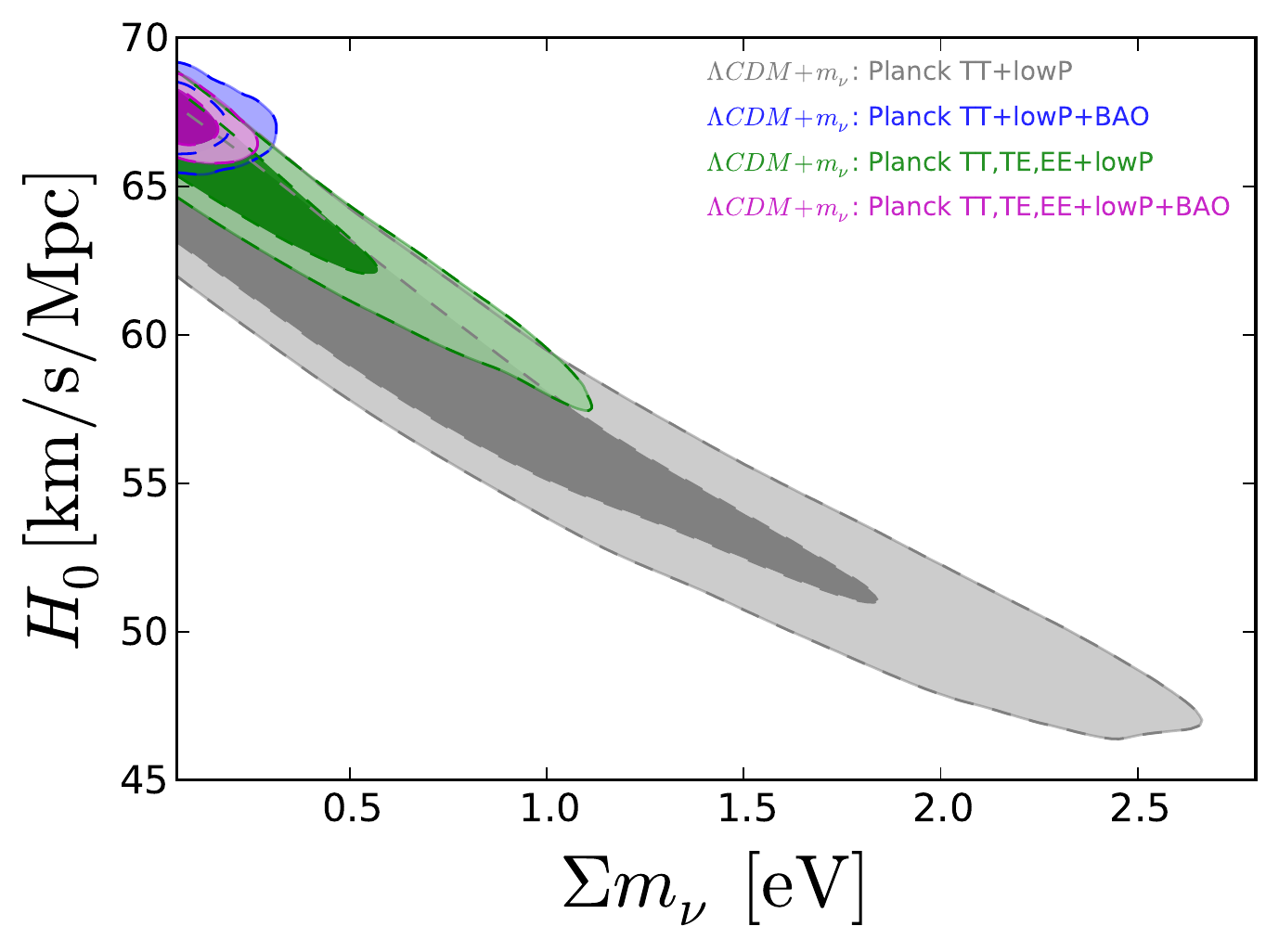}
 \caption{$68\%$ and $95\%$~CL allowed regions in the
 ($\mnu$, $H_0$) plane,
 using different combinations of datasets, within the \pchip PPS parameterization.}
\label{fig:fig1a_mnu}
\end{figure}

\begin{figure}[t]
\includegraphics[width=8.5cm]{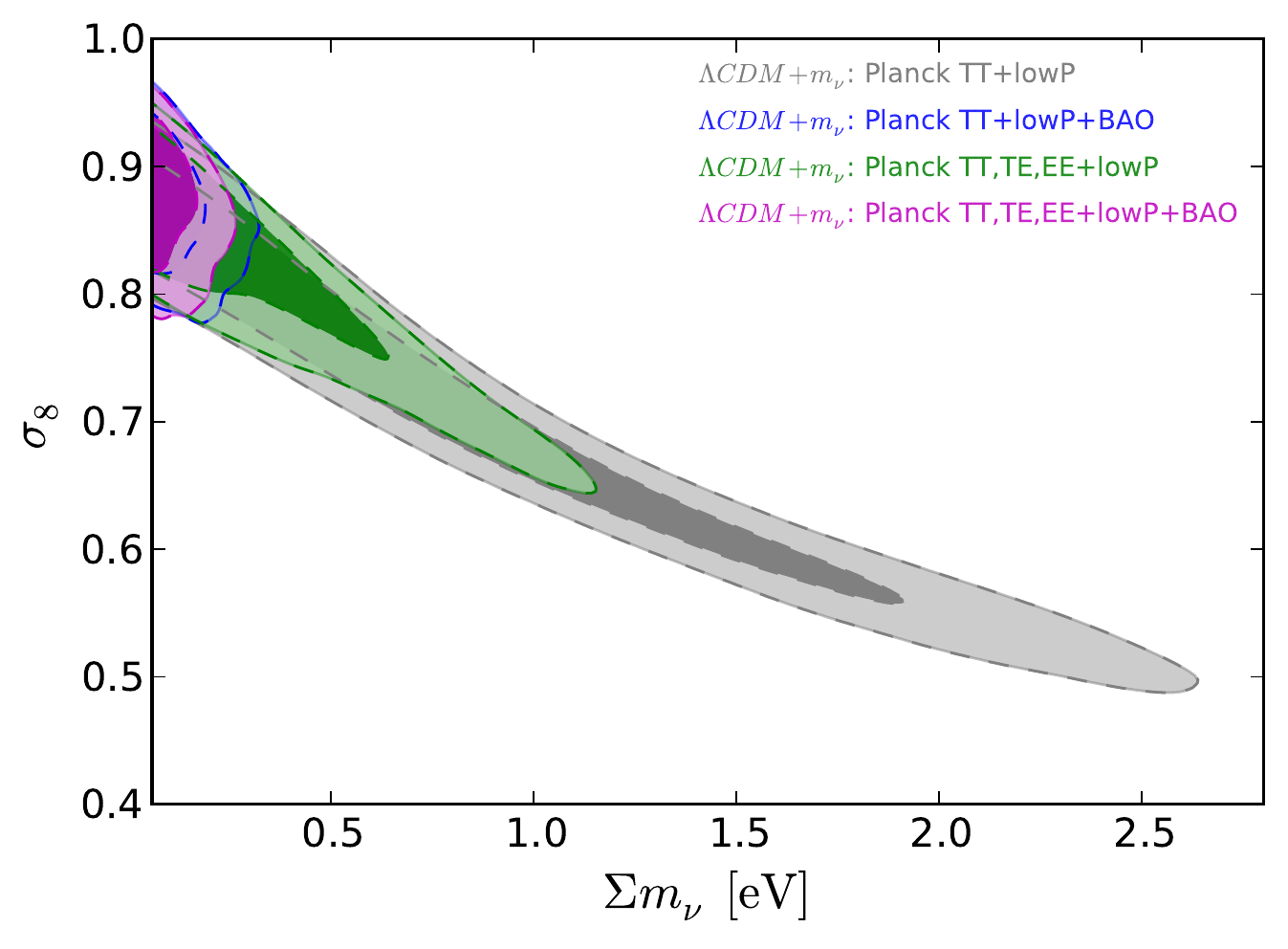}
 \caption{$68\%$ and $95\%$~CL allowed regions in the
 ($\mnu$, $\sigma_8$)
 plane,
 using different combinations of datasets, within the \pchip PPS parameterization.}
\label{fig:fig1b_mnu}
\end{figure}

It can be noticed that in the~\pchip PPS there is a shift in the Hubble constant toward lower values. This occurs because there exists a strong, well-known degeneracy between the neutrino mass and the Hubble constant, see Fig.~\ref{fig:fig1a_mnu}. In particular, considering CMB data only, a higher value of $\mnu$ will shift the location of the angular diameter distance to the last scattering surface, change that can be compensated with a smaller value of the Hubble constant $H_0$. The mean values of the clustering parameter $\sigma_8$ are also displaced by $\sim2\sigma$ (except for the BAO case) toward lower values in the~\pchip PPS approach with respect to the mean values obtained when using the power-law PPS, as can be noticed from Fig.~\ref{fig:fig1b_mnu}.
Concerning the $\psj{i}$ parameters,
the bounds on  $\psj{i}$ with $i\geq5$ are weaker with respect to the \lcdm case (see Tab.~\ref{tab:lcdm} in the Appendix), and only the combination of Planck TT+lowP data with the MPkW measurements provides an upper limit for the $\psj{12}$ (concretely, $\psj{12}<3.89$ at 95\% CL).

Also when considering the high-$\ell$ polarization measurements, the bounds on the sum of the neutrino masses are larger when using the \pchip parameterization with respect to the ones obtained with the power-law approach. However, these bounds are more stringent than those obtained using the Planck TT+lowP data only (see Tab.~\ref{tab:mnuTTTEEE} in the Appendix).
The reason for this improvement is due to the fact that the inclusion of the polarization measurements removes many of the degeneracies among the parameters. Concerning the CMB measurements only, we find an upper limit $\mnu<0.880$ eV at 95\% CL in the \pchip approach. The addition of the matter power spectrum measurements leads to a value of  $\mnu<0.458$ eV at 95\% CL in the \pchip parameterization, improving the Planck TT,TE,EE+lowP constraint by a factor of two. Notice that, as in the Planck TT+lowP results, the data combination that gives the most stringent constraints is the one involving the Planck TT,TE,EE+lowP and BAO datasets, since it provides a 95\% CL upper bound on  $\mnu$ of 0.218 eV in the \pchip PPS case.
Finally, when the lensing measurements are added, the constraint on the neutrino masses is shifted to a higher value (agreeing with previous findings from the Planck collaboration), being $\mnu<1.17$ eV at 95\% CL for the \pchip case. The degeneracies between $\mnu$ and $H_0$, $\sigma_8$, even if milder than those without high multipole polarization data, are still present (see Figs.~\ref{fig:fig1a_mnu} and \ref{fig:fig1b_mnu}). The constraints on the $\psj{i}$ parameters do not differ much from those obtained with the Planck TT+lowP data.

\section{Effective Number of Relativistic Species and Neutrino Masses}
\label{sec:mnu_nnu}
After having analyzed the constraints on $\neff$
and \mnu\ separately,  we study in this section their joint constraints in the context of the \lcdm + \neff\ + \mnu\ extended cosmological model, focusing mainly on the differences with the results presented in the two previous sections. 

\begin{figure}
\centering
\includegraphics[width=\columnwidth]{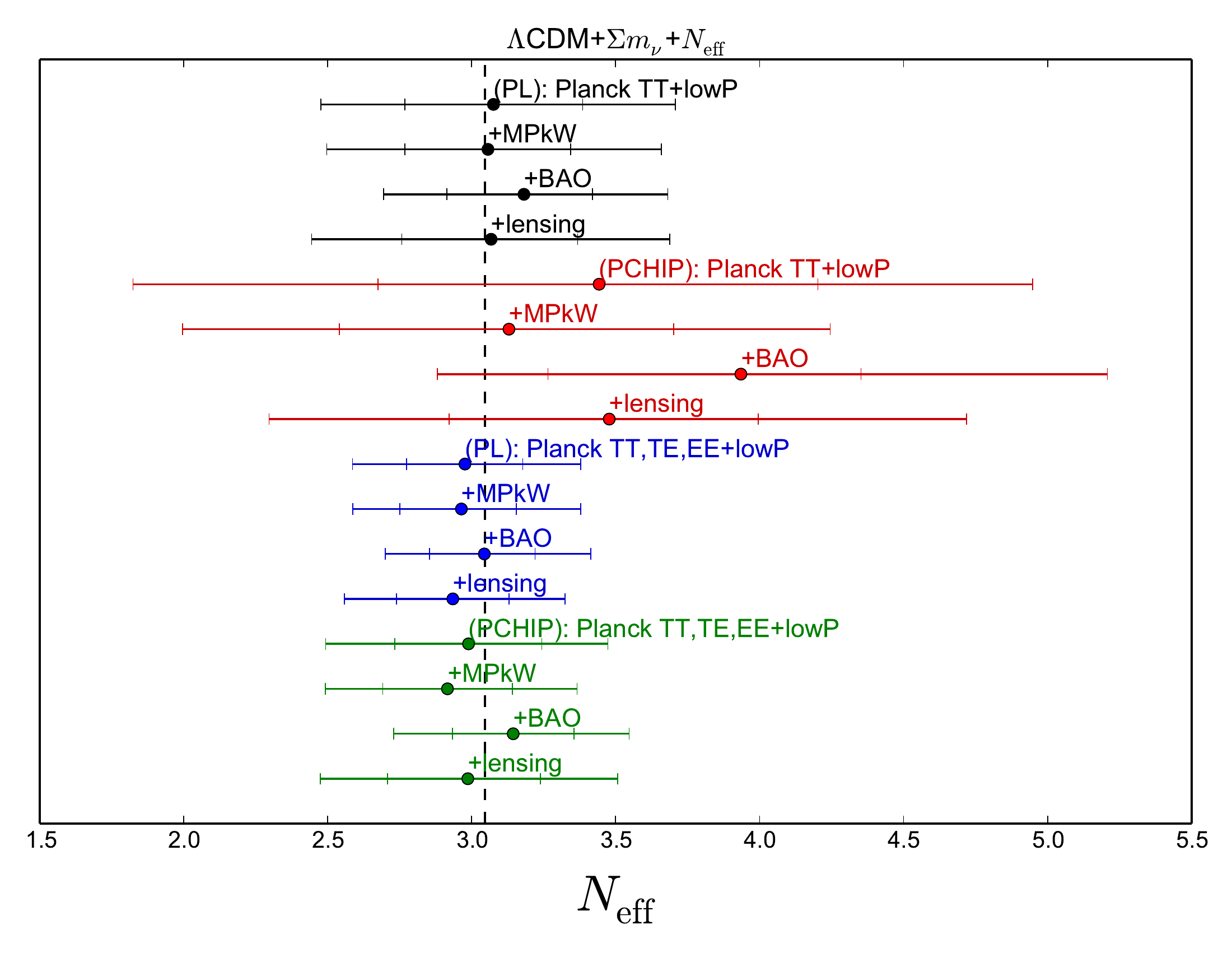}
\caption{As Fig.~\ref{fig:nnu_bars} but for the 
 \lcdm + \neff\ + \mnu\ model.}
\label{fig:mnu_nnu_bars_nnu}
\end{figure}
\begin{figure}
\centering
\includegraphics[width=\columnwidth]{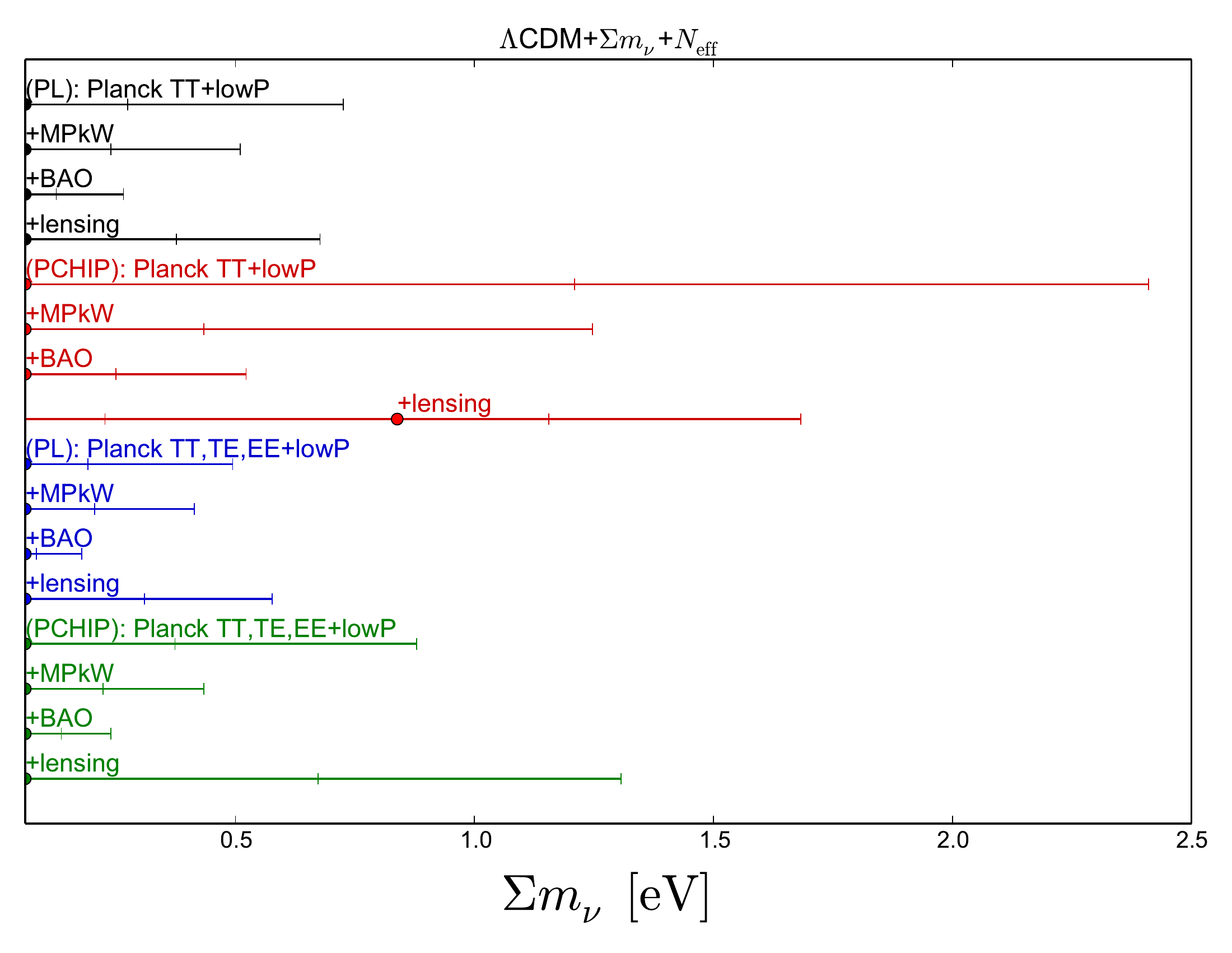}
\caption{As Fig.~\ref{fig:mnu_nnu_bars_nnu},
but for the $\mnu$ parameter.}
\label{fig:mnu_nnu_bars_mnu}
\end{figure}

The 68\% and 95\% CL constraints on \neff\ and \mnu\
are reported in Figs.~\ref{fig:mnu_nnu_bars_nnu} and
\ref{fig:mnu_nnu_bars_mnu} respectively, for different
dataset combinations and PPS choices, see also Tabs.~\ref{tab:mnu_nnu} and \ref{tab:mnu_nnu_pol} in the Appendix. 
Notice that the qualitative conclusions drawn in the previous sections do not change here.
The \pchip PPS parameterization still allows for a significant freedom in the values of \neff\ and \mnu, as these parameters have an impact on the CMB spectrum that can be easily mimicked by some variations in the PPS nodes. In particular, a significant degeneracy between \neff\ and the nodes \psj{6} to \psj{10} appears, in analogy to what happens in the \lcdm + \neff\ model (see Fig.~\ref{fig:nnu_corr} and the discussion in Sec.~\ref{sec:nnu}). At the same time, the strongest degeneracy involving the total neutrino mass appears between \mnu\ and \psj{5}. This corresponds to a rescaling of the PPS that compensates the change in the early ISW contribution driven by massive neutrinos (see Fig.~\ref{fig:mnu_corr} and the discussion in Sec.~\ref{sec:nu}). We do not show here the degeneracies with the \psj{i} nodes for the \lcdm + \neff\ + \mnu\ model, but we have verified that they are qualitatively similar to those depicted in Fig.~\ref{fig:nnu_corr} (Fig.~\ref{fig:mnu_corr}) for the \neff\ (\mnu) parameter. When considering CMB data only, the constraints are slightly loosened with respect to those obtained when the \neff\ and \mnu\ parameters are freely varied separately and not simultaneously.  When comparing the power-law PPS and the \pchip PPS models, we can notice that the variations of the neutrino parameters lead to several variations in other cosmological parameters. Such is the case of  the baryon and CDM densities and the angular scale of the peaks, which are shifted by a significant amount, as a consequence of the degeneracies
with both \neff\ and \mnu. As the effects of \neff\ and \mnu\ on the Hubble parameter $H_0$ and the clustering parameter $\sigma_8$ are opposite, we find an increased uncertainty in these parameters, being their allowed ranges significantly enlarged.

While the tightest neutrino mass bound arises from the Planck~TT+lowP+BAO dataset, the largest allowed mean value for \neff\ is also obtained for this very same data combination ($\neff=3.94_{-0.67}^{+0.42}$ at 68\% CL in the \pchip PPS analysis), showing the large degeneracy between \mnu\ and \neff. However, when including the Planck CMB lensing measurements, the 
trend is opposite to the one observed with the BAO dataset, with three times larger upper limits for \mnu\ and lower mean values for \neff. 
As stated before, the fact that lensing data prefer heavier neutrinos is well known
(see e.g.\ Sec.~\ref{sec:nu} and Refs.~\cite{Ade:2013zuv,Ade:2015xua}). Notice, from Fig.~\ref{fig:mnu_nnu_bars_mnu}, that the only 
combination which shows a preference for $\mnu>0.06$~eV\footnote{This value roughly corresponds to the lower limit allowed by oscillation
measurements if the total mass is distributed among the massive eigenstates according to the normal hierarchy scenario.} at 68\% CL
includes the lensing data ($\mnu=0.84^{+0.32}_{-0.62}$~eV, for the \pchip PPS).

When polarization measurements are added in the data analyses, we obtain a 95\% CL range of $2.5\lesssim\neff\lesssim3.5$, with very small differences in both the central values and allowed ranges for the several data combinations explored here, see Tab.~\ref{tab:mnu_nnu_pol} in the Appendix. As in the \lcdm + \neff\ model, the dataset including BAO data is the only one for which the mean value of \neff\ is larger than 3, while in all the other cases it lies between 2.9 and 3. Apart from these small differences, all the results are perfectly in agreement with the standard value 3.046 within the 68\% CL range.  Concerning the \mnu\ parameter, the results are also very similar to those obtained in the
\lcdm + \mnu\ model illustrated in Sec.~\ref{sec:nu}, with only very small differences in the exact numerical values of the 
derived bounds. The most constraining results are always obtained with the inclusion of 
BAO data, from which we obtain
$\mnu<0.18\; (0.24)$~eV when using the power-law (\pchip) PPS,
both really close to the values derived in the \lcdm + \mnu\ model.

For what concerns the remaining cosmological parameters,
the differences between the power-law PPS and the \pchip PPS results are much less significant when the polarization spectra are considered
in the analyses. We may notice that the predicted values of the Hubble parameter $H_0$
are lower than the CMB estimates in the \lcdm\ model, and consequently
they show an even stronger tension with local measurements of the Hubble constant. This is due to the negative correlation between $H_0$ and \mnu. On the other hand, the \lcdm + \neff\ + \mnu\ model 
predicts a $\sigma_8$ smaller than what is obtained in the \lcdm\ model
for most of the data combinations, partially reconciling
the CMB and the local estimates for this parameter.

The \pchip nodes in this extended model do not deviate significantly from
the expected values corresponding to the power-law PPS.
The small deviations driven by the degeneracies with the neutrino
parameters \mnu\ and \neff\ are canceled by the stringent bounds set by the polarization spectra, that break these degeneracies. Deviations from the power-law expectations are still visible at small wavemodes, corresponding to the dip at $\ell\simeq 20$ and to the small bump at $\ell\simeq40$ in the CMB temperature spectrum.

\section{Massive neutrinos and extra massive sterile neutrino species}\label{sec:meff}

Standard cosmology includes as hot thermal relics the three light, active neutrino flavors of the
Standard Model of elementary particles. However, the existence of extra hot relic components,
as dark radiation relics, sterile neutrino species and/or thermal axions is also possible.
In their presence, the cosmological neutrino
mass constraints will be changed. The existence of extra sub-eV massive sterile neutrino species is well motivated by the so-called short-baseline neutrino oscillation anomalies~\cite{Abazajian:2012ys,Kopp:2013vaa,Gonzalez-Garcia:2015qrr,Gariazzo:2015rra}.
These extra light species have an associated free streaming scale that will reduce the growth of matter fluctuations at small
scales. They also contribute to the effective number of relativistic degree of freedom (i.e. to $\neff$). 

We explore in this section the \lcdm scenario (in the two PPS parameterizations, power-law and \pchip) with three active
light massive neutrinos, plus one massive sterile neutrino
species characterized by an effective mass $\meff$, that is defined by
\begin{equation}
\meff=\left(\frac{T_s}{T_\nu}\right)^3 m_s=(\Delta \neff)^{3/4}m_s~,
\end{equation}
where $T_s$ ($T_\nu$) is the current temperature of the sterile (active)
neutrino species, $\Delta \neff\equiv\neff-3.046=(T_s/T_\nu)^3$ is the effective number of degrees of freedom associated to the massive sterile neutrino, and $m_s$ is its physical mass. For the numerical analyses we use the following
set of parameters to describe the model with a power-law PPS: 
\begin{equation}\label{eq:meff_parameter}
\{\Omega_b h^2,\Omega_c h^2, \theta, \tau, n_s, \log[10^{10}A_{s}], \mnu, \neff, \meff \} ~.
\end{equation}
When considering the \pchip PPS parameterization, $n_s$ and $A_s$ are replaced by the twelve parameters $\psj{i}$ (with $i=1,\ldots,12$).

\begin{figure}
\centering
\includegraphics[width=\columnwidth]{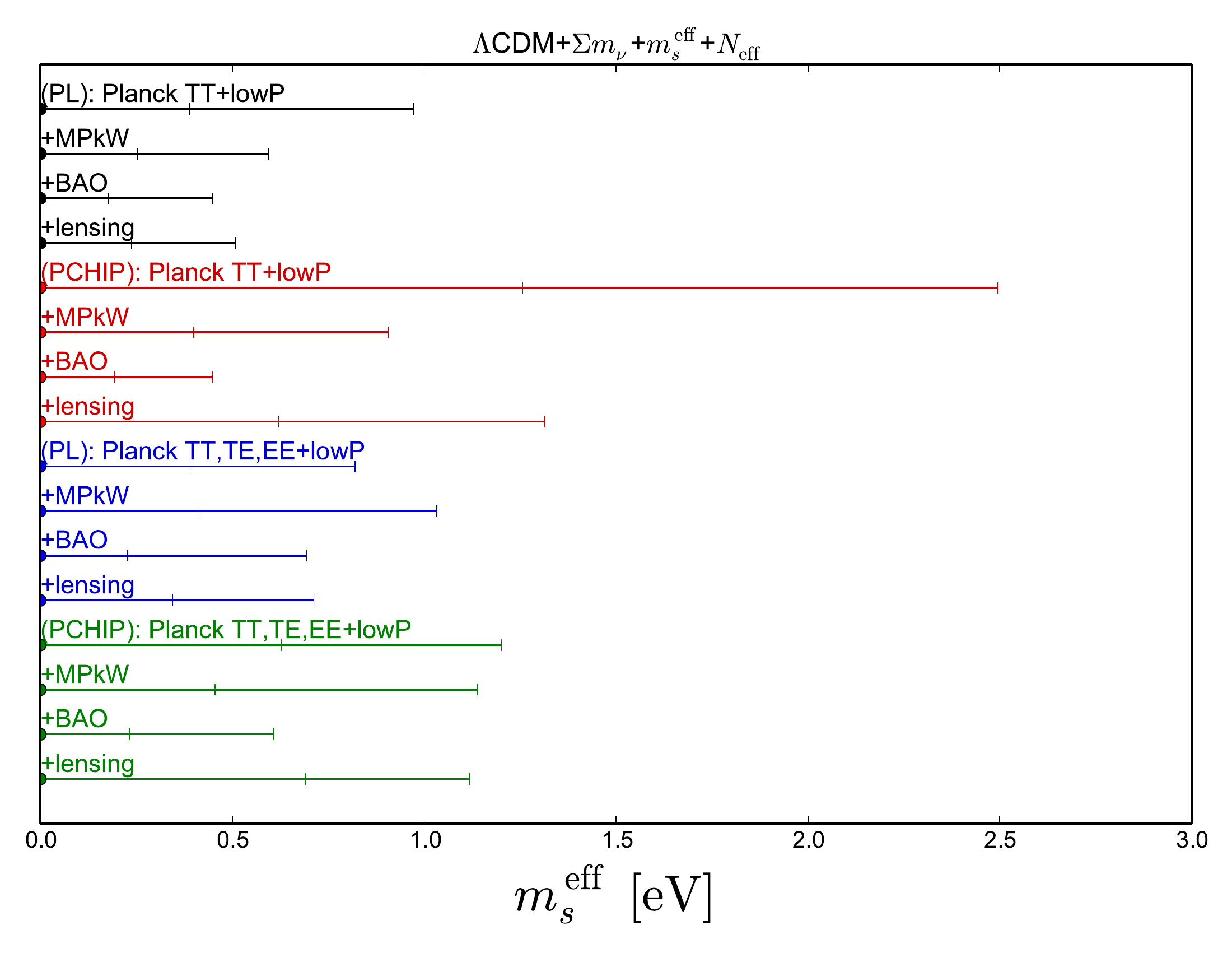}
\caption{68\% and 95\% CL constraints on $\meff$, obtained in the
 \lcdm + \neff\ +\mnu +\meff model.
 Different colors indicate
 Planck~TT+lowP with PL PPS (black),
 Planck~TT+lowP with PCHIP PPS (red),
 Planck~TT,TE,EE+lowP with PL PPS (blue) and
 Planck~TT,TE,EE+lowP with PCHIP PPS (green).
 For each color we plot 4 different datasets: from top to bottom, we have
 CMB only, CMB+MPkW, CMB+BAO and CMB+lensing.}
\label{fig:sterile_meff_bars}
\end{figure}

The 68\% and 95\% CL bounds on \meff\ obtained with different datasets and PPS combinations are summarized in Fig.~\ref{fig:sterile_meff_bars} and in Tabs.~\ref{tab:meffTT} and \ref{tab:meffTTTEEE} in the Appendix. Notice that, in general, the value of $\neff$ is larger than in the case in which the sterile neutrinos are considered massless (see Tab.~\ref{tab:nnu} in the Appendix).  As for the other extensions of the \lcdm\ model we studied, the bounds on  $\mnu$, $\neff$ and $\meff$ are weaker when considering the \pchip PPS with respect to the ones obtained within the power-law PPS canonical scenario. Notice that the bounds on $\meff$ are not very stringent. This is due to the correlation between \meff\ and \neff: sub-eV massive sterile neutrinos contribute to the matter energy density at recombination and therefore a larger value of \neff\ will be required to leave unchanged both the angular location and the height of the first acoustic peak of the CMB.

\begin{figure}[t]
\includegraphics[width=8.5cm]{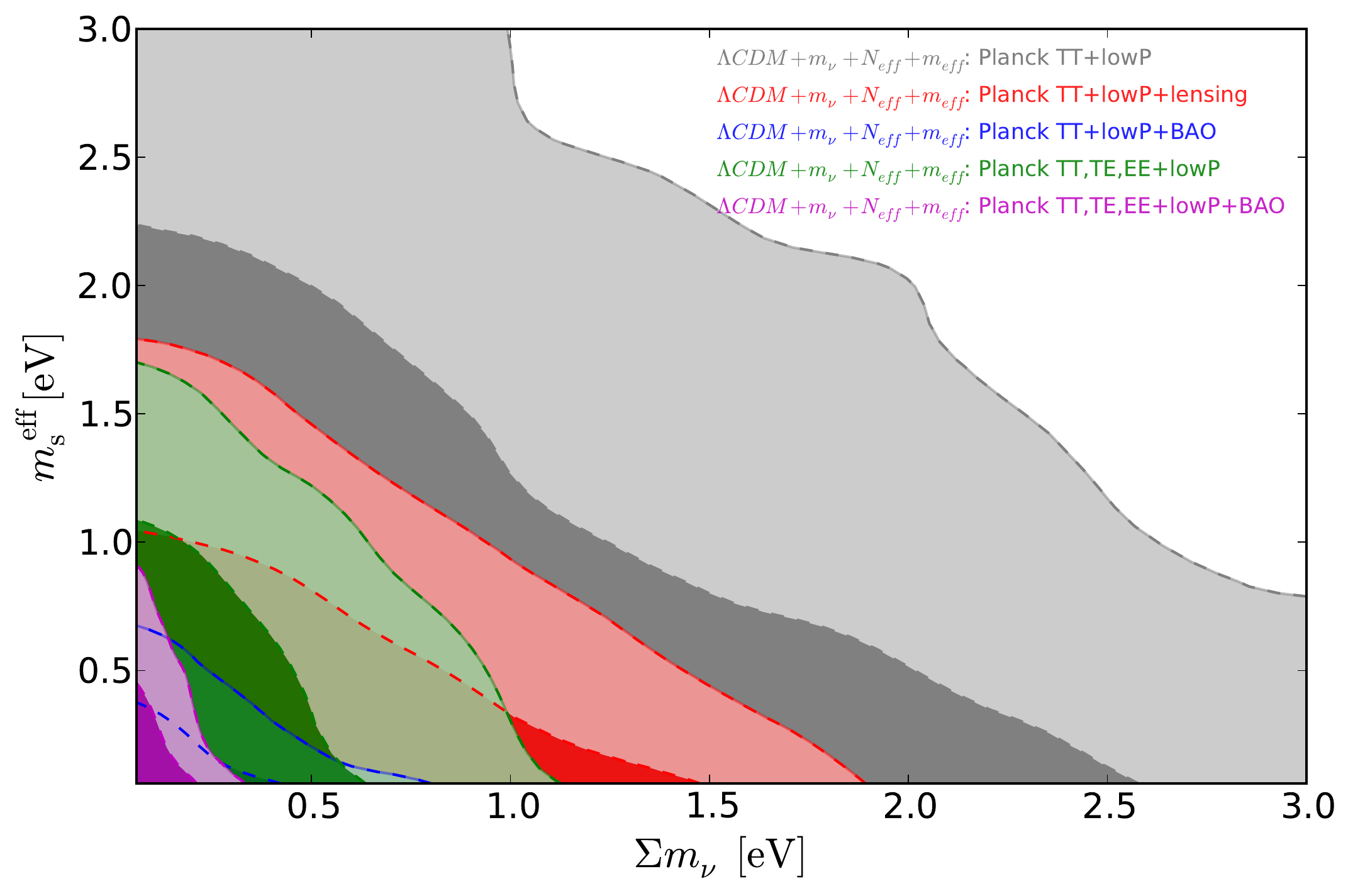}
 \caption{$68\%$ and $95\%$~CL allowed regions in the ($\mnu$, $\meff$) plane using the different combination of datasets, within the \pchip PPS parameterization.}
\label{fig:fig1}
\end{figure}

Figure~\ref{fig:fig1} illustrates the degeneracy between the active and the sterile neutrino masses.  Since both active and sterile
sub-eV massive neutrinos contribute to the matter energy density at decoupling, an increase of \meff\ can be compensated by lowering $\mnu$, in order to keep fixed the matter content of the universe. Notice that the most stringent  $95\%$ CL bounds on the three active and sterile neutrinos are obtained considering the BAO data in the two PPS cases.
In particular, we find $\mnu<0.481$ eV, $\meff<0.448$ eV for the \pchip parametrization and $\mnu<0.263$ eV, $\meff<0.449$ eV for the power-law approach. Furthermore, in general, when considering the \pchip parametrization, the mean value on the Hubble constant is smaller than the value obtained in the standard power-law PPS framework, due to the strong degeneracy between \mnu\ and $H_0$. The value of the clustering parameter $\sigma_8$ is reduced in the two PPS parameterizations when comparing to the  massless sterile neutrino case. This occurs because the sterile neutrino mass is another source of suppression of the large scale structure growth.

The inclusion of the polarization data improves notably the constraints on the cosmological parameters in the model with a \pchip parametrization. In particular, the neutrino constraints are stronger than
those obtained using only the temperature power spectrum at small angular scales. This effect is related to the fact that many degeneracies are reduced by the high multipole polarization measurements (as, for example, the one between $\mnu$ and $\tau$). Concerning the CMB measurements only, we find an upper limit on the three active and sterile neutrino masses of $\mnu<0.83$ eV and $\meff<1.20$ eV at 95\% CL, while for the effective number of relativistic degrees of freedom we obtain $\neff<3.67$ at 95\% CL, considering the \pchip PPS approach.  Also in this case the most stringent constraints are obtained when adding the BAO datasets to the Planck TT,TE,EE+lowP data. Finally, the addition of the lensing potential displaces both the active and sterile neutrino mass constraints to higher values. 

Concerning the $\psj{i}$ parameters, we can notice that considering the Planck TT,TE,EE+lowP+BAO datasets, the dip corresponding to the $\psj{3}$ node is reduced with respect to the other possible data combinations. We have an upper bound for the $\psj{12}$ node from all the data combinations except for the CMB+lensing dataset combination. In addition, as illustrated in Secs.~\ref{sec:nu} and \ref{sec:mnu_nnu}, a significant degeneracy between \neff\ and the nodes $\psj{8}$ to $\psj{10}$ and between \mnu\ and the nodes $\psj{5}$ and $\psj{6}$ is also present in this \lcdm extension.
Finally, because of the correlation between \mnu\ and \meff,  degeneracies between \meff\ and the nodes $\psj{5}$ and $\psj{6}$ will naturally appear.

\section{Thermal Axion}\label{sec:ax}

\begin{figure}
\centering
\includegraphics[width=\columnwidth]{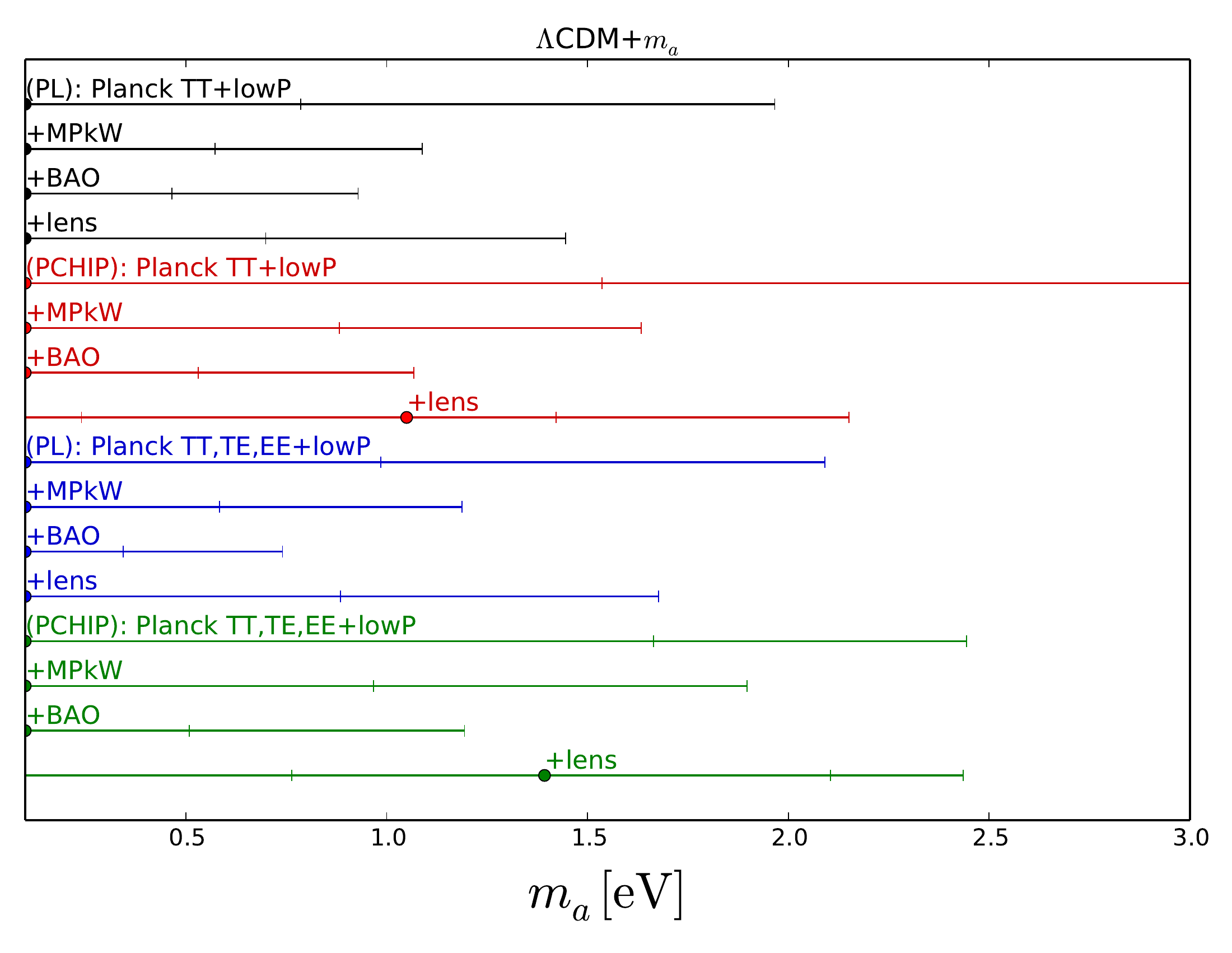}
\caption{As Fig.~\ref{fig:nnu_bars} but in the context of the
 \lcdm + $m_a$ model, focusing on the thermal axion mass $m_a$ parameter.}
 \label{ma_bars}
\end{figure}

\begin{figure*}
\centering
\includegraphics[width=0.9\textwidth]{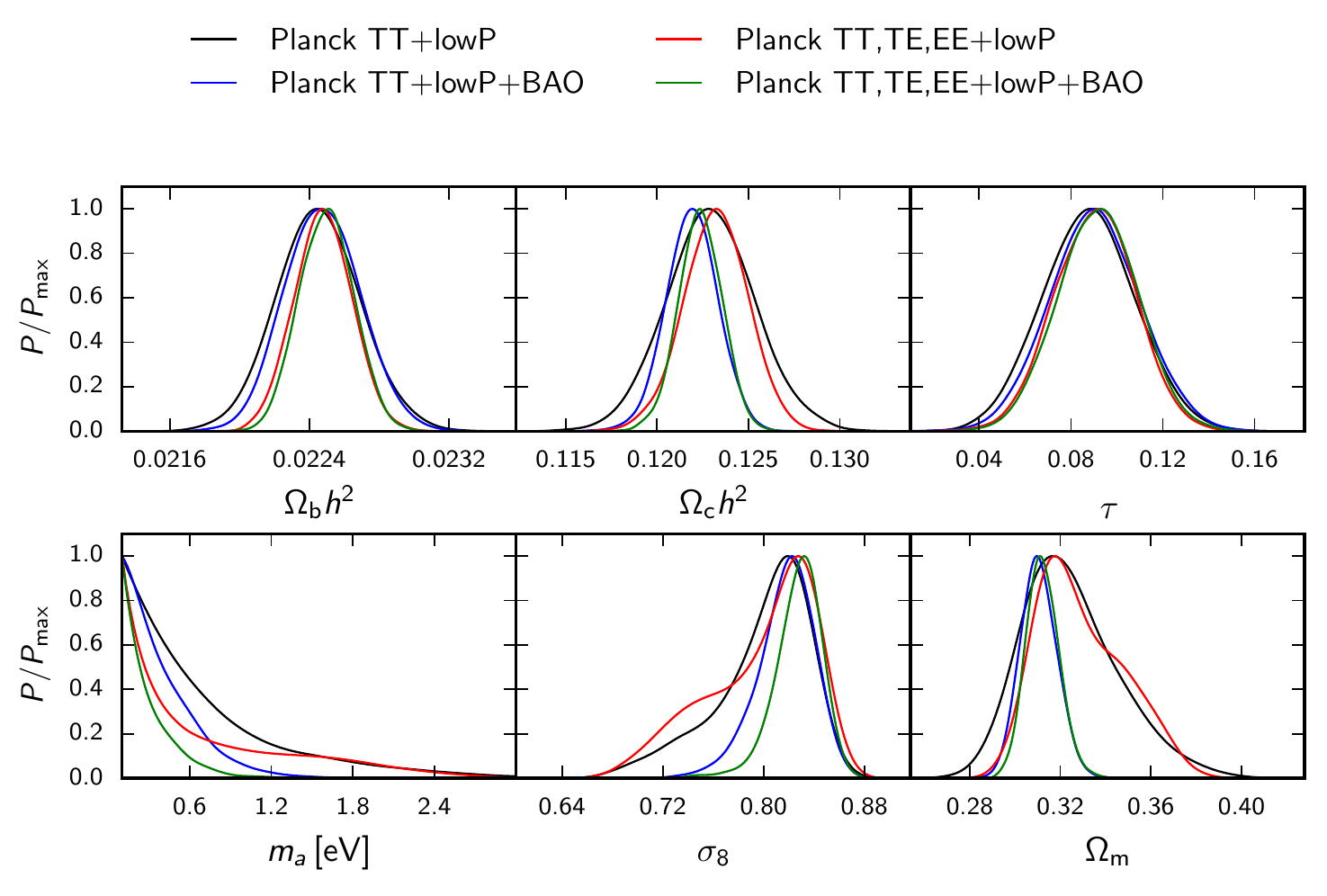}
\caption{One-dimensional posterior probability for some cosmological parameters for the combination of datasets labeled in the figure, for the power-law approach in the \lcdm + $m_a$ scenario.}\label{1dma}
\end{figure*}

\begin{figure}
\centering
\includegraphics[width=\columnwidth]{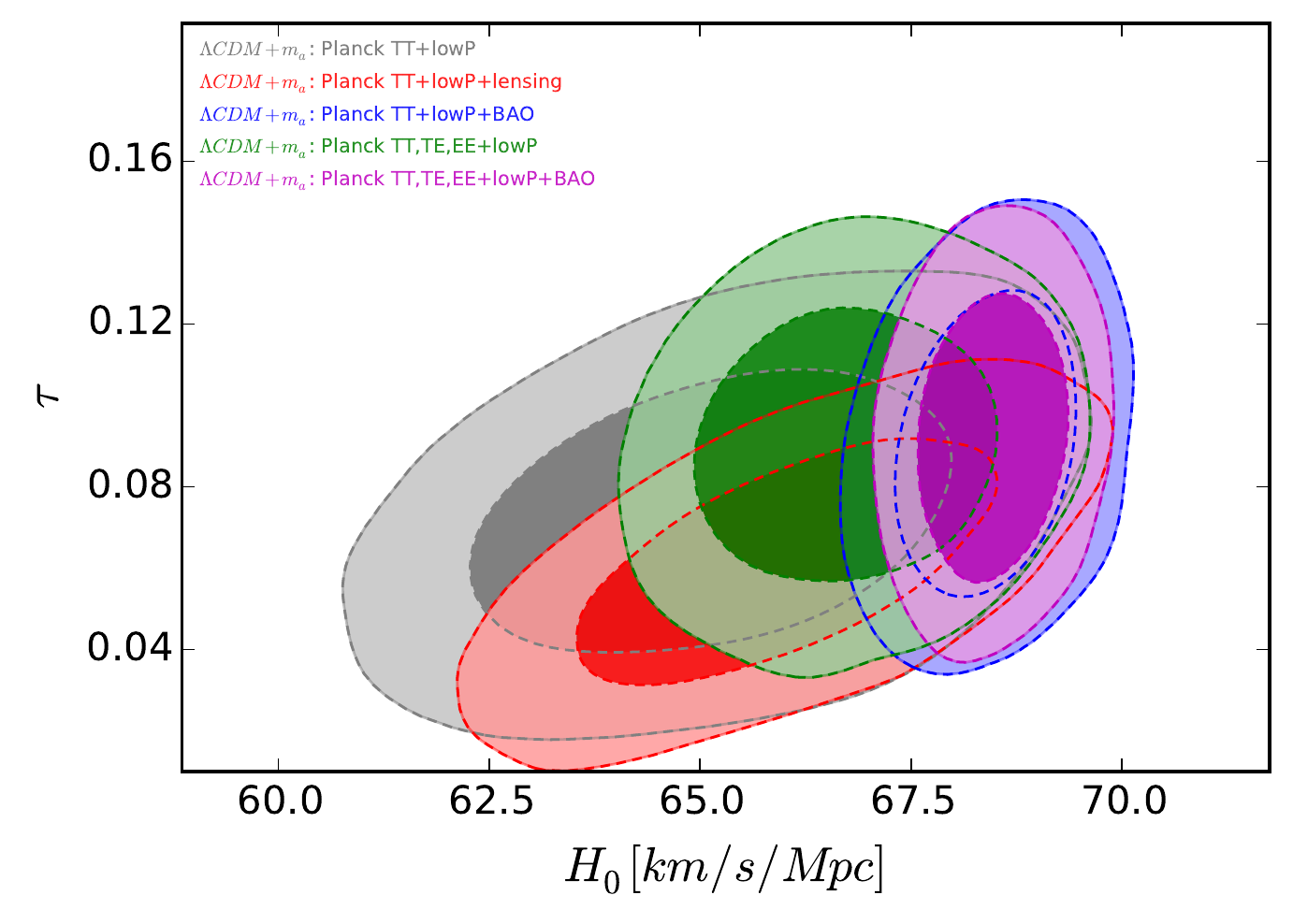}
\caption{$68\%$ and $95\%$~CL allowed regions in the ($H_0$, $\tau$) plane using the different combinations of datasets in the \lcdm + $m_a$ model  with a \pchip PPS.}\label{H0_tau}
\end{figure}

The axion field arises from the solution proposed by Peccei and Quinn~\cite{Peccei:1977hh,Peccei:1977ur,Weinberg:1977ma,Wilczek:1977pj} 
to solve the strong CP problem in Quantum Chromodynamics. They introduced a new global Peccei-Quinn symmetry $U(1)_{PQ}$ that, when spontaneously broken at an energy scale $f_a$, generates a Pseudo-Nambu-Goldstone boson, the axion particle. Depending on the production process in the early universe, thermal or non-thermal, the axion is a possible candidate for an extra hot thermal relic, together with the relic neutrino background, or for the cold dark matter component, respectively. In what follows, we shall focus on the thermal axion scenario. The axion coupling constant $f_a$ is related to the thermal axion mass via
\begin{equation}
m_a = \frac{f_\pi m_\pi}{  f_a  } \frac{\sqrt{R}}{1 + R}=
0.6\ {\rm eV}\ \frac{10^7\, {\rm GeV}}{f_a}~,
\end{equation}
with $R=0.553 \pm 0.043$, the up-to-down quark masses
ratio, and $f_\pi = 93$ MeV, the pion decay constant. Considering other values of $R$ within the range $0.38-0.58$ \cite{Agashe:2014kda} does not affect in a significant way this relationship \citep{Archidiacono:2015mda}. 

When the thermal axion is still a relativistic particle, it increases the effective number of relativistic degrees of freedom $\neff$, enhancing the amount of radiation in the early universe, see  Eq.~(\ref{eq:rhorad}). It is possible to compute the contribution 
of a thermal axion as an extra radiation component as:
\begin{equation}\label{eq:deltaneff}
\Delta \neff =\frac{4}{7}\left(\frac{3}{2}\frac{n_a}{n_\nu}\right)^{4/3}~,
\end{equation}
with $n_a$ the current axion number density and $n_\nu$ the present neutrino plus antineutrino number density per flavor. When the thermal axion becomes a non relativistic particle, it increases the amount of the hot dark matter density in the universe, contributing to the total mass-energy density of the universe. Thermal axions promote clustering only at large scales, suppressing the structure formation at scales smaller than their free-streaming scale,  once the axion is a non-relativistic particle. Several papers in the literature provide bounds on the thermal axion mass, see for example Refs.~\cite{Melchiorri:2007cd,Hannestad:2007dd,Hannestad:2008js,Hannestad:2010yi,Archidiacono:2013cha,Giusarma:2014zza, DiValentino:2015wba}. In this paper our purpose is to update the work done in Ref.~\cite{DiValentino:2015zta}, in light of the recent Planck 2015 temperature and polarization data \cite{Adam:2015rua}. Therefore, in what follows, we present up-to-date constraints on the thermal axion mass, relaxing the assumption of a power-law for the PPS of the scalar perturbations, assuming also the \pchip PPS scenario.

The bounds on the axion mass are relaxed in the \pchip PPS scenario, as illustrated in Fig.~\ref{ma_bars} (see also Tabs.~\ref{tab:maTT} and \ref{tab:maTTTEEE} in the Appendix). This effect is related to the relaxed bound we have on $N_{\textrm{eff}}$ when letting it free to vary in an extended \lcdm + $N_{\textrm{eff}}$ scenario. From the results presented in  Tab.~\ref{tab:nnu},  we find $N_{\textrm{eff}}=3.40 _{-1.43}^{+1.50}$ at $95\%$~CL for the \pchip PPS parameterization, implying that the \pchip formalism favours extra dark radiation, and therefore a higher axion mass will be allowed. As a consequence, we find that the axion mass is totally unconstrained using the Planck~TT+lowP data in the \pchip PPS approach. We instead find the bound $m_a<1.97$ eV at $95\%$~CL for the standard power-law case. 
The most stringent bounds arise when using the BAO data, since they are directly sensitive to the free-streaming nature of the thermal axion. While the MPkW measurements are also sensitive to this small scale structure suppression, BAO measurements  are able to constrain better the cold dark matter density $\Omega_c h^2$, strongly correlated with $m_a$. We find $m_a<0.93$ eV at $95\%$~CL in the standard case, and a slightly weaker constraint in the \pchip case, $m_a<1.07$ eV at $95\%$~CL. Finally, when considering the lensing dataset, we obtain $m_a<1.45$ eV at $95\%$~CL in the power-law PPS case, bound that is relaxed in the \pchip PPS, $m_a<2.15$ eV at $95\%$~CL. For this combination of datasets, a mild preference appears for an axion mass different from zero ($m_a=1.05_{-0.81}^{+0.37}$ at $68\%$~CL), only when considering the \pchip approach, as illustrated in Fig.~\ref{ma_bars}. This is probably due to the existing tension between the Planck lensing reconstruction data and the lensing effect, see Refs.~\cite{Ade:2015xua,DiValentino:2015ola}.

The weakening of the axion mass constraints in most of the data combinations obtained in the \pchip PPS scheme is responsible for the shift at more than 1$\sigma$ in the cold dark matter mass-energy density, due to the existing degeneracy between $m_a$ and $\Omega_c h^2$. Interestingly, this effect has also an impact on the Hubble  constant, shifting its mean value by about 2$\sigma$ towards lower values, similarly to the results obtained in the neutrino mass case.
Furthermore, a shift in the optical depth towards a lower mean value is also present when analyzing the \pchip PPS scenario. One can explain this shift via the existing degeneracies between $\tau$ and $H_0$ and between $\tau$ and $\Omega_c h^2$. Once BAO measurements are included in the data analyses, the degeneracies are however largely removed and there is no significant shift in the values of the $\Omega_c h^2$, $H_0$ and $\tau$ parameters within the \pchip PPS approach, when compared to their mean values in the power-law PPS.   Concerning the $\psj{i}$ parameters, we can observe also in this \lcdm + $m_a$ scenario a dip (corresponding to the $\psj{3}$ node) and a bump (corresponding to the $\psj{4}$ node), see Tabs.~\ref{tab:maTT} and \ref{tab:maTTTEEE} in the Appendix. These features are more significant for the case of CMB data only.

In general, the constraints arising from the addition of high-$\ell$ polarization measurements are slightly weaker than those previously 
obtained. The weakening of the axion mass is driven by the preference of Planck~TT,TE,EE+lowP for a lower value of $N_{\textrm{eff}}$, 
as pointed out before. As shown in Ref.~\cite{DiValentino:2015zta}, the additional contribution to $N_{\textrm{eff}}$ due 
to thermal axions is a steep function of the axion mass, at least for low thermal axion masses (i.e.\ below $\sim1$eV). 
The lower value of $N_{\textrm{eff}}$ preferred by small-scale polarization dramatically sharpens the posterior of $m_a$ at low mass 
(see Fig.~\ref{1dma}). At higher masses, axions contribute mostly as cold dark matter: the posterior distribution flattens and overlaps 
with the one resulting from Planck TT+lowP, since CMB polarization does not help in improving the constraints on $\Omega_m$ 
(notice the presence of a bump in the posterior distributions of $\Omega_m$ and $\sigma_8$ for Planck~TT,TE,EE+lowP). The mismatch in the values of $\Omega_m$ preferred by low and high thermal axion masses leads to a worsening in the constraints on $m_a$ with respect to the Planck TT+lowP scenario, since the volume of the posterior distribution is now mainly distributed at higher masses. When BAO data are considered, we get the tightest bounds on $m_a$. This is due to the fact that BAO measurements allow to constrain better $\Omega_m$, excluding the high mass axion region. In addition, the bump in both the $\Omega_m$ and $\sigma_8$ distributions disappears completely, due to the higher constraining power on the clustering parameter and the matter density. As can be noticed from Fig.~\ref{1dma}, the tail of the $m_a$ distribution is excluded when adding BAO measurements.

Furthermore, the thermal axion mass bounds are relaxed within the \pchip PPS formalism. In particular, concerning the CMB measurements only, $m_a<2.44$ eV at $95\%$~CL in the \pchip approach, compared to the bound $m_a<2.09$ eV at $95\%$~CL in the standard power-law PPS description. When adding the matter power spectrum measurements (MPkW) we find upper limits on the axion mass that are $m_a<1.19$ eV at $95\%$~CL in the power-law PPS and $m_a<1.90$ eV at $95\%$~CL in the \pchip parametrization. When considering the lensing dataset, we obtain $m_a<1.68$ eV at $95\%$~CL in the power-law PPS case, that is relaxed in the \pchip PPS, $m_a<2.44$ eV at $95\%$~CL. A mild preference for an axion mass different from zero appears from this particular data combination ($m_a=1.39_{-0.63}^{+0.71}$ at $68\%$~CL) only when considering a \pchip approach, see Fig.~\ref{ma_bars}.

It is important to note that, when high multipole polarization data is included, there is no shift induced neither in the mean value of the optical depth nor in the one corresponding to the cold dark matter energy density in the \pchip approach (with respect to the power-law case). 
High $\ell$ polarization data is extremely powerful in breaking degeneracies, as, for instance, the one existing between $\tau$ and $H_0$, as noticed from Fig.~\ref{H0_tau}.

\begin{figure}
\centering
\includegraphics[width=\columnwidth]{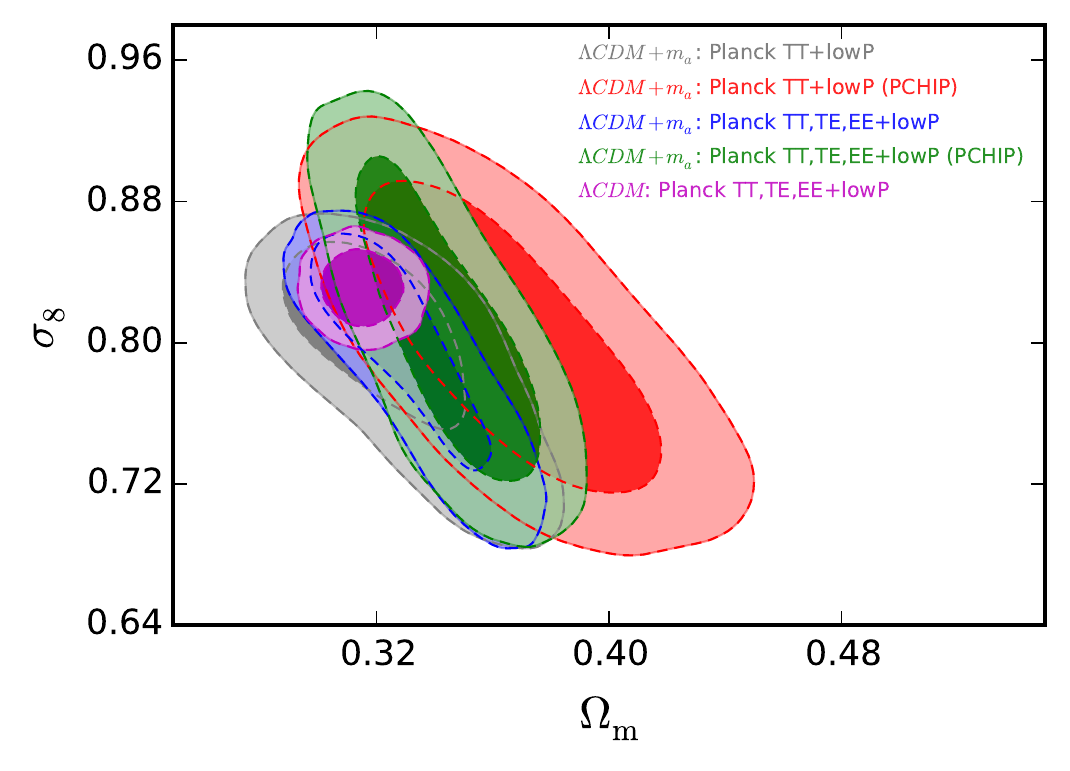}
\caption{$68\%$ and $95\%$~CL allowed regions in the ($\Omega_m$, $\sigma_8$) plane using the different combinations of datasets in the \lcdm + $m_a$ model.}\label{omsig8}
\end{figure}

 Interestingly, varying the thermal axion mass has a significant effect in the $\sigma_8-\Omega_m$ plane in both PPS approaches, see Fig.~\ref{omsig8}, weakening the bounds found for the \lcdm model and pushing the $\sigma_8$ ($\Omega_m$) parameter towards a lower (higher) value.  Concerning the $\psj{i}$ parameters, the bounds on the nodes remain unchanged after adding the high-$\ell$ polarization data (when they are compared to the Planck~TT+lowP baseline case). The significance of the dip and the bump are also very similar for  the different datasets.

We have also explored the case in which both massive neutrinos and a thermal axion contribute as possible hot dark matter candidates. Our results, not illustrated here, show that the thermal axion mass bounds are unchanged in the extended \lcdm + $m_a$ + $\mnu$ model with respect to the \lcdm + $m_a$ scenario, leading to almost identical  axion mass contraints. On the other hand, the presence of thermal axions tightens the neutrino mass bounds, as these two thermal relics behave as hot dark matter with a free-streaming nature. The most stringent bounds on both the axion mass and on the total neutrino mass arise, as usual, from the addition of BAO data. We find $m_a<1.18$ eV at $95\%$~CL and $\mnu<0.180$ eV at $95\%$~CL in the \pchip PPS when combining BAO with the Planck TT,TE,EE+lowP datasets. 

\section{Primordial Power Spectrum Results}
\label{sec:pps}
\begin{figure}
  \centering
  \includegraphics[width=0.99\columnwidth]{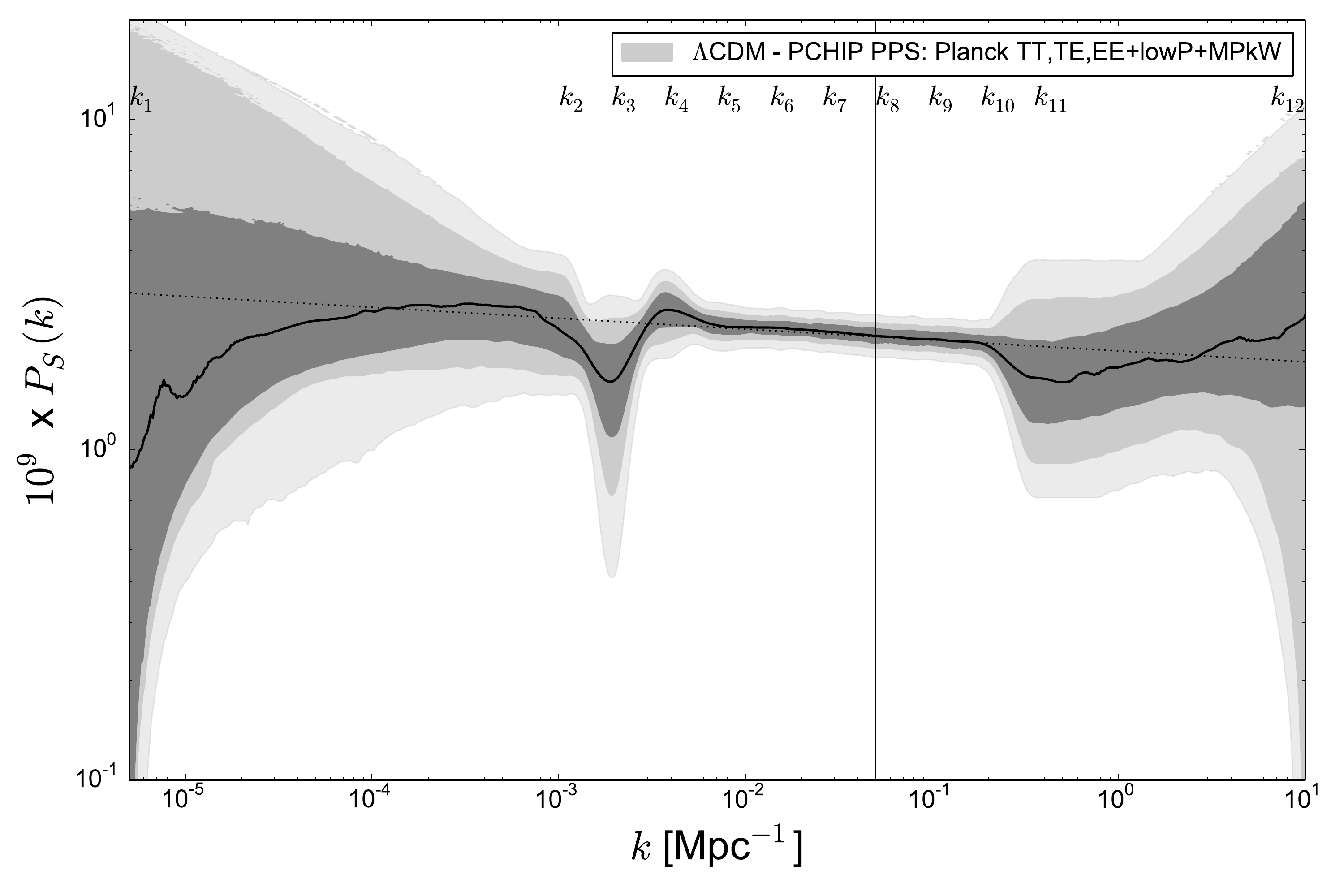}
  \caption{\label{fig:pps_planck_pol_mpkw} 
   Reconstruction of the marginalized best fit
    \pchip~PPS (solid line) with 68\%, 95\% and 99\% confidence bands
    as obtained in the \lcdm~model,
    with the ``'Planck~TT,TE,EE+lowP+MPkW'' dataset.
    The dotted line represents the power-law PPS
    corresponding to the Planck
    best fit \cite{Ade:2015xua}.}
\end{figure}

From the MCMC analyses presented in the previous sections we obtained 
constraints on the nodes used to parameterize the \pchip~PPS.
Using these information, we can
obtain a reconstruction of the spectrum shape for the different extensions
of the \lcdm~model.
Since the form of the reconstructed PPS
is similar for the different models,
we discuss now the common features of the \pchip~PPS
as obtained for the 
\lcdm~model.
We shall comment on the results for the dataset combinations shown in Tab.~\ref{tab:lcdm}:
Planck~TT+lowP, Planck~TT,TE,EE+lowP,
Planck~TT+lowP+MPkW  and Planck~TT,TE,EE+lowP+MPkW.
 Figure \ref{fig:pps_planck_pol_mpkw} illustrates the results for this last dataset combination. 
For all datasets the nodes $\psj{1}$ and $\psj{12}$
are badly constrained, due to the fact that these nodes are selected
to cover a wide range of wavemodes for computational reasons,
but there are no available data to constrain them directly.
Also the node $\psj{11}$ is not very well constrained
by the Planck temperature
data, however the bounds on $\psj{11}$ and $\psj{12}$
can be improved with the inclusion of the
high-multipole polarization data (TE,EE), leading to a significant improvement for $\psj{11}$.
The inclusion of the MPkW data allows to further tighten the constraints
on the last two nodes of the \pchip~PPS parameterization,
see Fig.~\ref{fig:pps_planck_pol_mpkw}. The impact of the polarization on the nodes located at high $k$ is smaller than
the one due to the addition of the matter power spectrum data, since the MPkW dataset provides
very strong constraints on the smallest angular scales.

The bounds on the nodes at small wavemodes ($\psj{1}$ to $\psj{4}$)
are almost insensitive to the inclusion of additional datasets
or to the change in the underlying cosmological model,
with only small variations inside the 1$\sigma$ range
between the different results.
The error bars on the nodes are larger in this part of the spectrum,
since it corresponds to low multipoles in the CMB power spectra,
where cosmic variance is larger.
In this part of the PPS we have the most evident deviations from the
simple power-law PPS.
The features are described by the node $\psj{3}$, 
for which the value corresponding to the power-law PPS is
approximately $2\sigma$ away from the reconstructed result,
and by the node $\psj{4}$, 
which is mildly discrepant with the power-law value ($1\sigma$ level).
These nodes describe the behavior of the CMB temperature spectrum
at low-$\ell$, where the observations of the Planck and WMAP experiments
show a lack of power at $\ell\simeq20$ and an excess of power
at $\ell\simeq40$.
The detection of these features is in agreement with several previous
studies
\cite{Shafieloo:2003gf,Nicholson:2009pi,Hazra:2013eva,Hazra:2014jwa,
Nicholson:2009zj,Hunt:2013bha,Hunt:2015iua,
Goswami:2013uja,Matsumiya:2001xj,Matsumiya:2002tx,
Kogo:2003yb,Kogo:2005qi,Nagata:2008tk,Ade:2015lrj,
Gariazzo:2014dla,DiValentino:2015zta}~\footnote{Since this behavior of the CMB spectrum at low multipoles
has been reported by analyses of both Planck and WMAP data,
it is unlikely that it is the consequence of some instrumental
systematics.
It is possible that this feature is simply the result of
a large statistical fluctuation in a region of the spectrum
where cosmic variance is very large.
On the other hand, the lack of power at a precise scale
can be the signal of some non-standard
inflationary mechanism that produced a non standard
spectrum for the initial scalar perturbations.
Future investigations will possibly clarify this aspect of the PPS.}.

The central part of the reconstructed PPS, from
\psj{5} to \psj{10}, is very well constrained by the data.
In this range of wavemodes, no deviations from the power-law PPS
are visible, thus confirming the validity of the assumption that
the PPS is almost scale-invariant for a wide range of wavemodes.
This is also the region where the PPS shape is more sensitive
to the changes in the
\lcdm~model caused by its extensions.

As we can see from the results presented in previous sections,
the constraints on the nodes \psj{5} to \psj{10} are different for each extension
of the \lcdm~model, in agreement with
the results obtained for $\ln[10^{10}A_s]$
and $n_s$ when considering the power-law PPS.
The value of the power-law PPS normalized to match the values of the
\pchip nodes can be calculated by means of
the relation $P_s(k)=A_s (k/k_*)^{n_s-1}/P_0$:
at each scale $k$, the value $P_s(k)$ is influenced both from $A_s$ and $n_s$,
since $P_0$ and $k_*$ are fixed.
In the various tables, when presenting the results on the power-law PPS,
we listed the values of the \pchip nodes that would correspond 
to the best-fitting $A_s$ and $n_s$,
to help in the comparison with the \pchip PPS constraints.
These values are calculated using Eq.~\eqref{eq:psjbf}.
In the range between $k\simeq 0.007$ and $k\simeq0.2$,
the constraints in the \pchip nodes
correspond, for most of the cases, to the values expected by the power-law PPS
analyses, within their allowed 1$\sigma$ range.
There are a few exceptions: for example,
in the \lcdm + \neff\ model and with the Planck~TT+lowP+BAO dataset,
the node \psj{10} deviates from the expected value corresponding
to the power-law PPS
by more than 1$\sigma$ (see Tab.~\ref{tab:nnu}).
This is a consequence of the large correlation and the large variability range
that this dataset allows for \neff.
A similar behavior appears in
the \lcdm + \neff + \mnu\ model (see Tab.~\ref{tab:mnu_nnu})
and in
the \lcdm + \mnu + \neff + $\meff$ model (Tab.~\ref{tab:meffTT}), 
for the same reasons.
The inclusion of polarization data at high-$\ell$,
limiting the range for \neff, 
does not allow for these deviations from the power-law PPS.

\begin{figure}
  \centering
  \includegraphics[width=0.99\columnwidth]{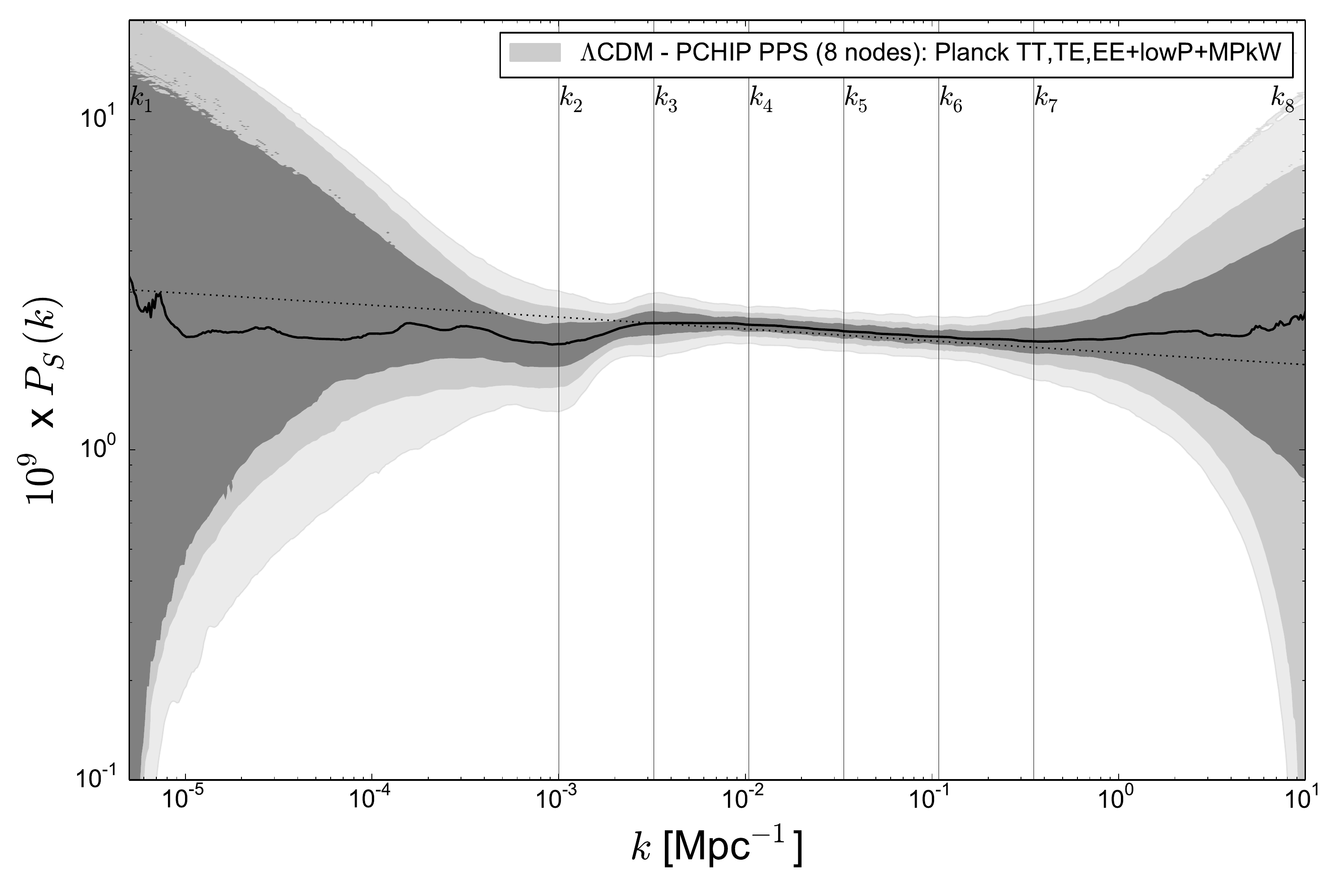}
  \caption{\label{fig:pps8n_planck_pol_mpkw} 
   Reconstruction of the marginalized best fit
    \pchip~PPS (solid line) with 68\%, 95\% and 99\% confidence bands
    as obtained in the \lcdm~model,
    with the ``'Planck~TT,TE,EE+lowP+MPkW'' dataset,
    using a \pchip\ parameterization with 8 nodes for the PPS.
    The dotted line represents the power-law PPS
    corresponding to the Planck
    best fit \cite{Ade:2015xua}.}
\end{figure}

It is interesting to study how the previous findings depend on the choice of the PPS parameterization. One could ask then how many nodes are needed to capture hints for unexplored effects, which could be due to unaccounted systematics, or, more interestingly, to new physics. A number of nodes larger than the one explored here (12 nodes) becomes unfeasible, as it would be extremely challenging computationally. However, lowering the number of nodes would be a very efficient solution for practical purposes, assuming the hints previously found are not totally diluted.
We have therefore checked this alternative scenario, using eight nodes, as described in Sec.~\ref{ssec:pps}.
The constraints on the PPS derived using this parameterization are reported
in Fig.~\ref{fig:pps8n_planck_pol_mpkw}, obtained considering the \lcdm\ model
and the Planck~TT,TE,EE+lowP+MPkW dataset.
The \pchip\ parameterization with only eight nodes is not able to catch the features that are observed at $k\simeq0.002\mpcinv$ with twelve nodes, since there are not enough nodes at the relevant wavemodes to describe the dip and the bump observed in the CMB spectrum.
Having less nodes, the PPS can describe less features,
it is more stable and the behavior at small and high $k$ can change.
We found a preference for higher values for the node in $k_1^{(8)}$ than the one in $k_1$,
as a consequence of the rules of the \pchip\ function for fixing the first derivatives in the nodes.
For the same reason, the constraints on the nodes in $k_{11}$ and $k_7^{(8)}$
are slightly different, with a smaller preferred value for the 8-nodes case.
We recall that the regions at extreme wavemodes, however, are not well constrained by the experimental data.
In the central region, where the CMB data are extremely precise, there is no difference between the two parameterizations. 
There is also no significant difference between the 8-nodes and the 12-nodes approaches when considering bounds on either the effective number of relativistic degrees of freedom $\neff$ or the total neutrino mass $\sum m_\nu$; in both cases the cosmological constraints on $\neff$ and $\sum m_\nu$ get very close to the expected ones in the power law PPS description after polarization measurements are included in the analyses, see the points obtained from the 8-nodes analysis case
in Figs.~\ref{fig:nnu_bars} and \ref{fig:mnu_bars}. 

In order to illustrate which PPS parameterization, among the three possible ones considered here, is preferred by current cosmological data, 
we compare, in the following, the  minimum $\chi^2$ resulting in each case from a fit to the Planck~TT,TE,EE+lowP+MPkW dataset. 
The minimum $\chi^2$ for the power-law, the \pchip model with 12 nodes and the \pchip scenario with 8 nodes is 13400, 13396 and 13392, 
respectively.  The difference between the minimum $\chi^2$ for the power-law approach and the \pchip model with 
12 nodes is $\Delta \chi^2=8$. These two models differ by $10$ parameters, which means that data prefer, 
albeit with a very poor statistical significance, the power-law PPS description. The same conclusion is reached when comparing 
instead the power-law and the \pchip scenario with 8 nodes. Therefore, though current cosmological data seems to prefer the power-law description, the statistical significance of this preference is still very mild, which may be sharpened by future measurements.



\section{Conclusions}\label{sec:conclusions}
The description of the cosmological model
may require a non-standard power-law
Primordial Power Spectrum (PPS) of scalar perturbations generated
during the inflationary phase at the beginning of the Universe.
Several analyses have considered the possible deviations from the PPS power-law exploiting both the WMAP and the Planck data measurements of the CMB temperature
power spectrum~\cite{Shafieloo:2003gf,Nicholson:2009pi,Hazra:2013eva,Hazra:2014jwa,
Nicholson:2009zj,Hunt:2013bha,Hunt:2015iua,
Goswami:2013uja,Matsumiya:2001xj,Matsumiya:2002tx,
Kogo:2003yb,Kogo:2005qi,Nagata:2008tk,Ade:2015lrj,
Gariazzo:2014dla,DiValentino:2015zta}.
Even if the significance for such deviations is small, it leaves some freedom for the PPS assumed form. 
Here we test the robustness of the cosmological bounds on several cosmological parameters when the PPS is allowed to have a model-independent shape, that we describe using
a \pchip\ function to interpolate a series of twelve or eight nodes \psj{j}.

We have explored the impact of a non-canonical PPS in several different extensions of the \lcdm\ model, varying the effective number of relativistic species, the masses of the active and the light sterile neutrinos, the neutrino perturbations and a thermal axion mass.

Concerning the effective number of degrees of freedom \neff,
we find that the results are in good agreement
with the standard value of 3.046,
if one assumes the standard power-law PPS.
Increasing \neff\ has the main effect of increasing the Silk damping of the CMB spectrum
at small scales and therefore it is easy change the PPS shape at that scales to compensate the increased damping.
This results in a strong degeneracy
between the relevant \pchip\ PPS nodes and \neff.
As a consequence of volume effects in the Bayesian analyses,
the constraints on \neff\ are significantly loosened. For some data combinations we obtain $\neff\simeq4.8$ allowed at 95\% CL.
However, the \pchip\ PPS nodes and \neff\ effects can not be compensated in the polarization spectra,
in particular in the case of the TE cross-correlation. This is the reason for which the inclusion of CMB polarization measurements
in the analyses allows to break the degeneracies and to restore the \neff\ bounds very close to 3.046 for all the data combinations, with 
$\neff>3.5$ excluded at more than 95\% CL for all the datasets.

In the minimal three active massive neutrinos scenario, the constraints on \mnu\ are relaxed with respect to the PPS power-law ones. This is due to the degeneracy between \mnu\ and the nodes $\psj{5}$ and $\psj{6}$,
that correspond to the scales at which the early Integrated Sachs-Wolfe (eISW)
effect contributes to the CMB spectrum. The tightest limit we find is $\mnu<0.218$ eV at 95\% CL from the combination of Planck~TT,TE,EE+lowP+BAO data.  The situation is not significantly changed when the effective number and
the neutrino masses are varied simultaneously, since the degeneracies with the PPS parameterization are different for these two neutrino parameters. Their constraints are slightly relaxed as a consequence of the increased parameter space.
The strongest bounds on \mnu\ arise from the Planck~TT,TE,EE+lowP+BAO
dataset, for which we obtain
$\mnu<0.18\,\mathrm{eV}$ (power-law PPS) and 
$\mnu<0.24\,\mathrm{eV}$ (\pchip\ PPS).
For \neff, all the data combinations including the Planck CMB polarization measurements give similar constraints, summarized as  $2.5\lesssim\neff\lesssim3.5$ at 95\% CL.

In the case in which we consider both massive neutrinos and massive sterile neutrino species, the bounds on \mnu, \meff\, and \neff\ are weaker for the \pchip approach when compared to the standard  power-law PPS parameterization. This occurs because there exist degeneracies between these parameters and some nodes of the \pchip\ PPS.
The most stringent constraints on the active and the sterile neutrino parameters are obtained  from the combination of Planck~TT,TE,EE+lowP+BAO, for which we find $\mnu<0.199$ eV, $\meff<0.69$ eV and $\neff< 3.41$ at 95\% CL in the power-law PPS  description and $\mnu<0.219$ eV, $\meff<0.61$ eV and $\neff< 3.53$ at 95\% CL within the \pchip PPS parameterization. 

Regarding the thermal axion scenario, we notice that the axion mass bounds are largely relaxed when using the \pchip approach.
When including the small scale CMB polarization we find a further weakening of the axion mass constraints: the reduced volume of the posterior distribution for small axion masses ($m_a$) is translated into a broadening of the marginalized constraints towards higher values for $m_a$.
The strongest bound we find on the thermal axion mass within the \pchip approach is $m_a<1.07$ eV at 95\% CL when considering the Planck TT+lowP+BAO data combination (while, in the power-law scenario, $m_a<0.74$ eV at 95\% CL). Finally, when including massive neutrinos in addition to the thermal axions, we find that, while the bounds on the thermal axion mass are unaffected, the constraints on the total neutrino mass are tighter than those obtained without thermal axions. The strongest bounds we find for the thermal axion mass and the total neutrino mass in the \pchip approach are $m_a<1.03$ eV at 95\% CL and $\mnu<0.180$ eV at 95\% CL, when considering the Planck TT+lowP+BAO and Planck TT,TE,EE+lowP+BAO dataset combinations, respectively.
In the power-law PPS scenario the strongest bounds are $m_a<0.76$ eV at 95\% CL and $\mnu<0.159$ eV at 95\% CL, obtained for the Planck TT,TE,EE+lowP+BAO dataset.

In summary, we have shown that degeneracies among the parameters involved in the \lcdm model (and its possible extensions) and the PPS shape arise when considering CMB temperature power spectrum measurements only. Fortunately, these degeneracies  disappear with the inclusion of high-$\ell$ polarization data. This is due to the fact that all these cosmological parameters influence the TT, TE and EE spectra in different ways. This confirms the robustness of both the \lcdm\ model and
the simplest inflationary models, that predict a power-law PPS that successfully explains the observations at small scales. The large scale fluctuations of the CMB spectrum, however, seem to point towards
something new in the scenarios that describe inflation.
It must be clarified whether these features  are indicating a more complicated inflationary mechanism 
or are instead statistical fluctuations of the CMB temperature
anisotropies.

Furthermore, we have as well verified that current data, albeit showing a very mild preference for the power-law scenario, is far from robustly discarding the \pchip parameterization, and, therefore, future cosmological measurements are mandatory to sharpen the PPS profile. 

\section*{Acknowledgements}

The work of S.G. and M.G. was supported by the Theoretical Astroparticle Physics research
Grant No. 2012CPPYP7 under the Program PRIN 2012 funded by the
Ministero dell'Istruzione, Universit\`a e della Ricerca (MIUR). M.G. also acknowledges support by the Vetenskapsr\aa det (Swedish Research Council).
This work has been done within the Labex ILP (reference ANR-10-LABX-63) part of the Idex SUPER, and received financial state aid managed by the Agence Nationale de la Recherche, as part of the programme Investissements d'avenir under the reference ANR-11-IDEX-0004-02.
E.D.V. acknowledges the support of the European Research Council via the Grant  number 267117 (DARK, P.I. Joseph Silk).
O.M. is supported by PROMETEO II/2014/050, by the Spanish Grant FPA2014--57816-P of the MINECO, by the MINECO Grant SEV-2014-0398 and by PITN-GA-2011-289442-INVISIBLES.


%

\onecolumngrid
\appendix
\section{Tables}
\label{ap:ap1}
\begin{table*}[!ht]
\scalebox{0.8}{
\begin{tabular}{|c||c| c||c|c||c|c||c|c|}
\hline
Parameter	&\multicolumn{2}{c||}{Planck TT+lowP}	&\multicolumn{2}{c||}{Planck TT,TE,EE+lowP}	&\multicolumn{2}{c||}{Planck TT+lowP}	&\multicolumn{2}{c|}{Planck TT,TE,EE+lowP}\\
	&\multicolumn{2}{c||}{}	&\multicolumn{2}{c||}{}	&\multicolumn{2}{c||}{+MPkW}	&\multicolumn{2}{c|}{+MPkW}\\
\hline\hline
$\Omega_b h^2$                         & $0.02222\,^{+0.00045}_{-0.00043}$ & $0.02175\,^{+0.00077}_{-0.00076}$ & $0.02225\,^{+0.00032}_{-0.00030}$ & $0.02215\,^{+0.00038}_{-0.00037}$ & $0.02221\,^{+0.00044}_{-0.00045}$ & $0.02190\,^{+0.00072}_{-0.00070}$ & $0.02223\pm0.00031$ & $0.02214\,^{+0.00035}_{-0.00036}$	\\
$\Omega_c h^2$                         & $0.1197\,^{+0.0043}_{-0.0042}$    & $0.1231\,^{+0.0061}_{-0.0059}$    & $0.1198\pm0.0029$    & $0.1209\,^{+0.0035}_{-0.0034}$    & $0.1198\pm0.0039$    & $0.1223\,^{+0.0056}_{-0.0053}$    & $0.1200\,^{+0.0028}_{-0.0027}$    & $0.1210\pm0.0033$   	\\
$100\theta$                            & $1.0409\pm0.0009$    & $1.0405\pm0.0011$    & $1.0408\pm0.0006$    & $1.0407\pm0.0006$    & $1.0409\,^{+0.0009}_{-0.0010}$    & $1.0406\pm0.0010$    & $1.0408\pm0.0006$    & $1.0407\pm0.0006$   	\\
$\tau$                                 & $0.078\,^{+0.038}_{-0.036}$       & $0.073\,^{+0.044}_{-0.042}$       & $0.079\pm0.034$       & $0.082\pm0.040$       & $0.075\,^{+0.038}_{-0.039}$       & $0.076\,^{+0.048}_{-0.046}$       & $0.076\,^{+0.034}_{-0.033}$       & $0.083\,^{+0.038}_{-0.037}$      	\\
$n_S$                                  & $0.966\pm0.012$       & --                                & $0.964\pm0.010$       & --                                & $0.965\pm0.011$       & --                                & $0.964\pm0.009$       & --	\\
$\ln[10^{10}A_s]$                      & $3.089\,^{+0.072}_{-0.069}$       & --                                & $3.094\pm0.066$       & --                                & $3.084\,^{+0.073}_{-0.074}$       & --                                & $3.087\,^{+0.066}_{-0.065}$       & --	\\
$H_0\,\mathrm{[km\,s^{-1}\,Mpc^{-1}]}$ & $67.3\,^{+1.9}_{-1.8}$            & $65.7\pm2.7$            & $67.3\pm1.3$            & $66.8\pm1.5$            & $67.3\,^{+1.7}_{-1.8}$            & $66.1\pm2.5$            & $67.2\pm1.2$            & $66.7\,^{+1.5}_{-1.4}$           	\\
$\sigma_8$                             & $0.83\pm0.03$          & $0.87\pm0.06$          & $0.83\pm0.03$          & $0.88\,^{+0.05}_{-0.06}$          & $0.83\pm0.03$          & $0.84\,^{+0.04}_{-0.03}$          & $0.83\pm0.03$          & $0.83\pm0.03$         	\\
\hline
$\psj{1}$                              & $\equiv 1.365$                    & $<7.93$                           & $\equiv 1.397$                    & $<7.69$                           & $\equiv 1.371$                    & $<7.90$                           & $\equiv 1.388$                    & $<7.68$         	\\
$\psj{2}$                              & $\equiv 1.140$                    & $1.15\,^{+0.38}_{-0.35}$          & $\equiv 1.155$                    & $1.14\,^{+0.39}_{-0.36}$          & $\equiv 1.139$                    & $1.14\,^{+0.39}_{-0.36}$          & $\equiv 1.147$                    & $1.14\,^{+0.38}_{-0.36}$         	\\
$\psj{3}$                              & $\equiv 1.115$                    & $0.73\,^{+0.39}_{-0.37}$          & $\equiv 1.128$                    & $0.71\,^{+0.38}_{-0.35}$          & $\equiv 1.113$                    & $0.73\,^{+0.39}_{-0.38}$          & $\equiv 1.120$                    & $0.72\,^{+0.38}_{-0.37}$         	\\
$\psj{4}$                              & $\equiv 1.091$                    & $1.19\,^{+0.26}_{-0.25}$          & $\equiv 1.102$                    & $1.22\,^{+0.23}_{-0.22}$          & $\equiv 1.088$                    & $1.19\pm0.25$          & $\equiv 1.094$                    & $1.22\pm0.22$         	\\
$\psj{5}$                              & $\equiv 1.067$                    & $1.07\pm0.11$          & $\equiv 1.076$                    & $1.08\,^{+0.11}_{-0.10}$          & $\equiv 1.063$                    & $1.07\,^{+0.12}_{-0.11}$          & $\equiv 1.069$                    & $1.08\pm0.10$         	\\
$\psj{6}$                              & $\equiv 1.043$                    & $1.06\,^{+0.09}_{-0.08}$          & $\equiv 1.051$                    & $1.07\,^{+0.08}_{-0.08}$          & $\equiv 1.040$                    & $1.06\pm0.09$          & $\equiv 1.044$                    & $1.07\,^{+0.08}_{-0.07}$         	\\
$\psj{7}$                              & $\equiv 1.021$                    & $1.04\,^{+0.09}_{-0.08}$          & $\equiv 1.027$                    & $1.04\pm0.08$          & $\equiv 1.016$                    & $1.03\pm0.09$          & $\equiv 1.020$                    & $1.04\,^{+0.08}_{-0.07}$         	\\
$\psj{8}$                              & $\equiv 0.998$                    & $0.99\,^{+0.09}_{-0.08}$          & $\equiv 1.003$                    & $1.01\pm0.08$          & $\equiv 0.993$                    & $1.00\pm0.09$          & $\equiv 0.996$                    & $1.01\,^{+0.08}_{-0.07}$         	\\
$\psj{9}$                              & $\equiv 0.976$                    & $0.97\,^{+0.09}_{-0.08}$          & $\equiv 0.980$                    & $0.99\,^{+0.08}_{-0.07}$          & $\equiv 0.971$                    & $0.98\pm0.09$          & $\equiv 0.973$                    & $0.99\,^{+0.08}_{-0.07}$         	\\
$\psj{10}$                             & $\equiv 0.955$                    & $0.97\,^{+0.10}_{-0.09}$          & $\equiv 0.957$                    & $0.98\pm0.09$          & $\equiv 0.949$                    & $0.95\pm0.09$          & $\equiv 0.951$                    & $0.96\pm0.08$         	\\
$\psj{11}$                             & $\equiv 0.934$                    & $<4.03$                           & $\equiv 0.935$                    & $2.44\,^{+2.00}_{-2.37}$          & $\equiv 0.928$                    & $0.82\,^{+0.45}_{-0.38}$          & $\equiv 0.929$                    & $0.81\,^{+0.45}_{-0.38}$         	\\
$\psj{12}$                             & $\equiv 0.833$                    & nb                                & $\equiv 0.829$                    & nb                                & $\equiv 0.825$                    & $<3.93$                           & $\equiv 0.823$                    & $<3.44$         	\\
\hline
\end{tabular}}
\caption{\textbf{The \texorpdfstring{\lcdm}{LambdaCDM} Model} Constraints on the cosmological parameters from 
the Planck TT+lowP and 
Planck TT,TE,EE+lowP datasets, and also in combination with the matter power spectrum
shape measurements from WiggleZ (MPkW),
in the \lcdm\ model (\emph{nb} refers to \emph{no bound}).
For each combination,
we report the limits obtained for the two parameterizations
of the primordial power spectrum, namely the power-law model
(first column)
and the polynomial expansion (second column of each data combination).
Limits are at 95\% CL around the mean value
of the posterior distribution.
For each dataset, in the case of the power-law model,
the values of \psj{i} are computed according to
Eq.~\eqref{eq:psjbf}.
}\label{tab:lcdm}
\end{table*}

\begin{table*}[!ht]
\scalebox{0.8}{
\begin{tabular}{|c||c| c||c|c||c|c||c|c|}
\hline
Parameter	
	&\multicolumn{2}{c||}{Planck TT+lowP}
	&\multicolumn{2}{c||}{Planck TT+lowP+MPkW}
	&\multicolumn{2}{c||}{Planck TT+lowP+BAO}
	&\multicolumn{2}{c| }{Planck TT+lowP+lensing}\\
\hline\hline
$\Omega_b h^2$                         & $0.02230\,^{+0.00075}_{-0.00071}$ & $0.02189\,^{+0.00107}_{-0.00105}$ & $0.02221\,^{+0.00066}_{-0.00063}$ & $0.02186\,^{+0.00081}_{-0.00082}$ & $0.02233\pm0.00047$ & $0.02205\,^{+0.00060}_{-0.00057}$ & $0.02232\,^{+0.00074}_{-0.00069}$ & $0.02198\,^{+0.00093}_{-0.00091}$\\
$\Omega_c h^2$                         & $0.1205\,^{+0.0081}_{-0.0077}$    & $0.1272\,^{+0.0189}_{-0.0182}$    & $0.1198\,^{+0.0077}_{-0.0073}$    & $0.1226\,^{+0.0148}_{-0.0141}$    & $0.1207\,^{+0.0077}_{-0.0074}$    & $0.1294\,^{+0.0153}_{-0.0146}$    & $0.1195\,^{+0.0079}_{-0.0073}$    & $0.1287\,^{+0.0169}_{-0.0161}$   \\
$100\theta$                            & $1.0408\pm0.0011$    & $1.0402\,^{+0.0019}_{-0.0018}$    & $1.0409\pm0.0011$    & $1.0406\,^{+0.0017}_{-0.0016}$    & $1.0408\pm0.0011$    & $1.0400\,^{+0.0015}_{-0.0014}$    & $1.0410\pm0.0011$    & $1.0401\,^{+0.0017}_{-0.0015}$   \\
$\tau$                                 & $0.080\,^{+0.044}_{-0.042}$       & $0.076\,^{+0.050}_{-0.047}$       & $0.075\,^{+0.040}_{-0.039}$       & $0.075\,^{+0.048}_{-0.043}$       & $0.082\,^{+0.035}_{-0.036}$       & $0.079\,^{+0.046}_{-0.041}$       & $0.069\,^{+0.040}_{-0.038}$       & $0.066\,^{+0.042}_{-0.038}$      \\
$\neff$                                & $3.13\,^{+0.64}_{-0.63}$          & $3.40\,^{+1.50}_{-1.43}$          & $3.05\,^{+0.58}_{-0.54}$          & $3.06\,^{+1.04}_{-1.00}$          & $3.15\,^{+0.47}_{-0.44}$          & $3.63\,^{+0.91}_{-0.80}$          & $3.13\,^{+0.62}_{-0.61}$          & $3.62\,^{+1.31}_{-1.19}$         \\
$n_S$                                  & $0.969\,^{+0.032}_{-0.030}$       & --                                & $0.965\,^{+0.027}_{-0.026}$       & --                                & $0.971\,^{+0.018}_{-0.017}$       & --                                & $0.971\,^{+0.030}_{-0.028}$       & --                               \\
$\ln[10^{10}A_s]$                      & $3.096\,^{+0.095}_{-0.089}$       & --                                & $3.083\,^{+0.085}_{-0.084}$       & --                                & $3.100\,^{+0.074}_{-0.075}$       & --                                & $3.070\,^{+0.085}_{-0.079}$       & --                               \\
$H_0\,\mathrm{[km\,s^{-1}\,Mpc^{-1}]}$ & $68.0\,^{+5.7}_{-5.6}$            & $68.2\,^{+11.4}_{-11.1}$          & $67.3\,^{+4.8}_{-4.6}$            & $66.0\,^{+7.4}_{-7.2}$            & $68.3\,^{+3.0}_{-2.9}$            & $70.2\,^{+4.6}_{-4.2}$            & $68.5\,^{+5.6}_{-5.3}$            & $70.2\,^{+9.4}_{-8.8}$           \\
$\sigma_8$                             & $0.83\,^{+0.05}_{-0.04}$          & $0.88\,^{+0.10}_{-0.09}$          & $0.83\pm0.04$          & $0.84\pm0.06$          & $0.84\pm0.04$          & $0.90\pm0.08$          & $0.82\pm0.04$          & $0.88\pm0.08$         \\ \hline
$\psj{1}$                              & $\equiv 1.337$                    & $<7.96$                           & $\equiv 1.369$                    & $<7.97$                           & $\equiv 1.318$                    & $<8.06$                           & $\equiv 1.279$                    & $<7.87$                          \\
$\psj{2}$                              & $\equiv 1.135$                    & $1.14\,^{+0.40}_{-0.37}$          & $\equiv 1.138$                    & $1.14\,^{+0.39}_{-0.36}$          & $\equiv 1.130$                    & $1.14\,^{+0.41}_{-0.38}$          & $\equiv 1.097$                    & $1.14\,^{+0.39}_{-0.37}$         \\
$\psj{3}$                              & $\equiv 1.112$                    & $0.73\,^{+0.41}_{-0.38}$          & $\equiv 1.112$                    & $0.73\,^{+0.40}_{-0.37}$          & $\equiv 1.109$                    & $0.72\,^{+0.41}_{-0.38}$          & $\equiv 1.076$                    & $0.70\,^{+0.39}_{-0.37}$         \\
$\psj{4}$                              & $\equiv 1.090$                    & $1.20\,^{+0.27}_{-0.25}$          & $\equiv 1.087$                    & $1.19\pm0.25$          & $\equiv 1.088$                    & $1.20\,^{+0.27}_{-0.26}$          & $\equiv 1.056$                    & $1.18\,^{+0.26}_{-0.25}$         \\
$\psj{5}$                              & $\equiv 1.068$                    & $1.07\,^{+0.13}_{-0.12}$          & $\equiv 1.062$                    & $1.07\pm0.11$          & $\equiv 1.068$                    & $1.06\pm0.12$          & $\equiv 1.036$                    & $1.04\pm0.10$         \\
$\psj{6}$                              & $\equiv 1.047$                    & $1.06\,^{+0.10}_{-0.09}$          & $\equiv 1.038$                    & $1.06\,^{+0.09}_{-0.08}$          & $\equiv 1.048$                    & $1.06\pm0.09$          & $\equiv 1.017$                    & $1.03\,^{+0.07}_{-0.06}$         \\
$\psj{7}$                              & $\equiv 1.026$                    & $1.05\,^{+0.10}_{-0.09}$          & $\equiv 1.015$                    & $1.03\,^{+0.09}_{-0.08}$          & $\equiv 1.028$                    & $1.05\,^{+0.09}_{-0.08}$          & $\equiv 0.998$                    & $1.02\,^{+0.08}_{-0.07}$         \\
$\psj{8}$                              & $\equiv 1.005$                    & $1.00\,^{+0.11}_{-0.10}$          & $\equiv 0.992$                    & $1.00\,^{+0.10}_{-0.09}$          & $\equiv 1.009$                    & $1.02\pm0.09$          & $\equiv 0.979$                    & $0.99\pm0.09$         \\
$\psj{9}$                              & $\equiv 0.985$                    & $1.00\,^{+0.14}_{-0.13}$          & $\equiv 0.970$                    & $0.97\,^{+0.11}_{-0.10}$          & $\equiv 0.990$                    & $1.02\pm0.09$          & $\equiv 0.961$                    & $0.99\,^{+0.12}_{-0.11}$         \\
$\psj{10}$                             & $\equiv 0.965$                    & $1.01\,^{+0.20}_{-0.19}$          & $\equiv 0.948$                    & $0.95\,^{+0.15}_{-0.14}$          & $\equiv 0.972$                    & $1.05\pm0.12$          & $\equiv 0.943$                    & $1.02\pm0.17$         \\
$\psj{11}$                             & $\equiv 0.946$                    & $<3.78$                           & $\equiv 0.927$                    & $0.85\,^{+0.58}_{-0.45}$          & $\equiv 0.954$                    & $<3.83$                           & $\equiv 0.925$                    & $<3.55$                          \\
$\psj{12}$                             & $\equiv 0.853$                    & nb                                & $\equiv 0.824$                    & $<4.24$                           & $\equiv 0.865$                    & nb                                & $\equiv 0.840$                    & nb                               \\
\hline
\end{tabular}}
\caption{\textbf{Effective Number of Relativistic Species} Constraints on cosmological parameters from 
the Planck TT+lowP dataset
alone and in combination with the matter power spectrum
shape measurements from WiggleZ (MPkW),
the BAO data and the lensing constraints from Planck,
in the \lcdm + \neff\ model (\emph{nb} refers to \emph{no bound}).
For each combination,
we report the limits obtained for the two parameterizations
of the primordial power spectrum, namely the power-law model
(first column)
and the polynomial expansion (second column of each pair).
Limits are at 95\% CL around the mean value
of the posterior distribution.
For each dataset, in the case of power-law model,
the values of \psj{i} are computed according to
Eq.~\eqref{eq:psjbf}.
}
\label{tab:nnu}
\end{table*}

\begin{table*}[!ht]
\scalebox{0.78}{
\begin{tabular}{|c||c| c||c|c||c|c||c|c|}
\hline
Parameter	&\multicolumn{2}{c||}{Planck TT,TE,EE+lowP}	&\multicolumn{2}{c||}{Planck TT,TE,EE+lowP}	&\multicolumn{2}{c||}{Planck TT,TE,EE+lowP}	&\multicolumn{2}{c|}{Planck TT,TE,EE+lowP}\\
	&\multicolumn{2}{c||}{}	&\multicolumn{2}{c||}{+MPkW}	&\multicolumn{2}{c||}{+BAO}	&\multicolumn{2}{c|}{+lensing}\\
\hline\hline
$\Omega_b h^2$                         & $0.02220\pm0.00048$ & $0.02206\,^{+0.00054}_{-0.00055}$ & $0.02214\,^{+0.00047}_{-0.00046}$ & $0.02203\pm0.00049$ & $0.02229\pm0.00038$ & $0.02226\,^{+0.00041}_{-0.00040}$ & $0.02216\,^{+0.00045}_{-0.00046}$ & $0.02204\,^{+0.00055}_{-0.00053}$\\
$\Omega_c h^2$                         & $0.1191\,^{+0.0062}_{-0.0061}$    & $0.1197\,^{+0.0072}_{-0.0071}$    & $0.1186\,^{+0.0062}_{-0.0061}$    & $0.1191\,^{+0.0070}_{-0.0067}$    & $0.1192\,^{+0.0060}_{-0.0059}$    & $0.1203\,^{+0.0067}_{-0.0068}$    & $0.1178\,^{+0.0058}_{-0.0057}$    & $0.1184\,^{+0.0069}_{-0.0067}$   \\
$100\theta$                            & $1.0409\pm0.0009$    & $1.0408\,^{+0.0010}_{-0.0009}$    & $1.0409\pm0.0009$    & $1.0409\pm0.0009$    & $1.0409\,^{+0.0009}_{-0.0008}$    & $1.0407\pm0.0009$    & $1.0410\,^{+0.0009}_{-0.0008}$    & $1.0410\,^{+0.0010}_{-0.0009}$   \\
$\tau$                                 & $0.077\pm0.035$       & $0.081\,^{+0.040}_{-0.039}$       & $0.073\,^{+0.036}_{-0.035}$       & $0.080\,^{+0.039}_{-0.037}$       & $0.082\pm0.032$       & $0.087\pm0.040$       & $0.060\pm0.028$       & $0.064\,^{+0.034}_{-0.032}$      \\
$\neff$                                & $2.99\,^{+0.41}_{-0.39}$          & $2.96\,^{+0.49}_{-0.48}$          & $2.95\,^{+0.41}_{-0.39}$          & $2.91\,^{+0.46}_{-0.43}$          & $3.04\pm0.35$          & $3.09\pm0.40$          & $2.94\pm0.38$          & $2.92\,^{+0.48}_{-0.46}$         \\
$n_S$                                  & $0.962\pm0.019$       & --                                & $0.960\pm0.019$       & --                                & $0.966\pm0.015$       & --                                & $0.961\,^{+0.019}_{-0.018}$       & --                               \\
$\ln[10^{10}A_s]$                      & $3.088\pm0.074$       & --                                & $3.078\,^{+0.075}_{-0.072}$       & --                                & $3.098\,^{+0.067}_{-0.069}$       & --                                & $3.049\,^{+0.058}_{-0.056}$       & --                               \\
$H_0\,\mathrm{[km\,s^{-1}\,Mpc^{-1}]}$ & $66.8\,^{+3.2}_{-3.1}$            & $66.1\,^{+3.9}_{-3.8}$            & $66.5\pm3.1$            & $65.8\,^{+3.6}_{-3.4}$            & $67.5\pm2.4$            & $67.6\,^{+2.6}_{-2.5}$            & $66.7\pm3.0$            & $66.2\,^{+3.9}_{-3.7}$           \\
$\sigma_8$                             & $0.83\,^{+0.04}_{-0.03}$          & $0.87\pm0.07$          & $0.82\,^{+0.04}_{-0.03}$          & $0.83\pm0.04$          & $0.83\pm0.03$          & $0.88\,^{+0.06}_{-0.08}$          & $0.81\,^{+0.03}_{-0.02}$          & $0.86\pm0.06$         \\ \hline
$\psj{1}$                              & $\equiv 1.415$                    & $<7.62$                           & $\equiv 1.427$                    & $<7.79$                           & $\equiv 1.377$                    & $<7.27$                           & $\equiv 1.373$                    & $<8.15$                          \\
$\psj{2}$                              & $\equiv 1.157$                    & $1.14\,^{+0.38}_{-0.35}$          & $\equiv 1.154$                    & $1.14\,^{+0.38}_{-0.35}$          & $\equiv 1.150$                    & $1.14\,^{+0.38}_{-0.36}$          & $\equiv 1.117$                    & $1.14\,^{+0.38}_{-0.35}$         \\
$\psj{3}$                              & $\equiv 1.128$                    & $0.72\,^{+0.37}_{-0.34}$          & $\equiv 1.125$                    & $0.72\,^{+0.37}_{-0.35}$          & $\equiv 1.125$                    & $0.73\,^{+0.38}_{-0.37}$          & $\equiv 1.089$                    & $0.68\,^{+0.36}_{-0.34}$         \\
$\psj{4}$                              & $\equiv 1.101$                    & $1.22\pm0.22$          & $\equiv 1.096$                    & $1.22\pm0.22$          & $\equiv 1.100$                    & $1.23\,^{+0.22}_{-0.21}$          & $\equiv 1.062$                    & $1.20\pm0.21$         \\
$\psj{5}$                              & $\equiv 1.074$                    & $1.08\pm0.10$          & $\equiv 1.068$                    & $1.08\,^{+0.10}_{-0.09}$          & $\equiv 1.076$                    & $1.09\,^{+0.11}_{-0.10}$          & $\equiv 1.035$                    & $1.05\,^{+0.09}_{-0.08}$         \\
$\psj{6}$                              & $\equiv 1.048$                    & $1.06\pm0.08$          & $\equiv 1.040$                    & $1.06\,^{+0.08}_{-0.07}$          & $\equiv 1.053$                    & $1.07\,^{+0.09}_{-0.08}$          & $\equiv 1.009$                    & $1.03\,^{+0.07}_{-0.06}$         \\
$\psj{7}$                              & $\equiv 1.022$                    & $1.04\pm0.08$          & $\equiv 1.013$                    & $1.04\,^{+0.08}_{-0.07}$          & $\equiv 1.030$                    & $1.05\,^{+0.09}_{-0.08}$          & $\equiv 0.984$                    & $1.00\pm0.06$         \\
$\psj{8}$                              & $\equiv 0.997$                    & $1.00\,^{+0.09}_{-0.08}$          & $\equiv 0.987$                    & $1.00\,^{+0.08}_{-0.07}$          & $\equiv 1.007$                    & $1.02\,^{+0.09}_{-0.08}$          & $\equiv 0.959$                    & $0.97\pm0.07$         \\
$\psj{9}$                              & $\equiv 0.973$                    & $0.98\,^{+0.09}_{-0.08}$          & $\equiv 0.962$                    & $0.98\,^{+0.09}_{-0.08}$          & $\equiv 0.985$                    & $1.00\,^{+0.09}_{-0.08}$          & $\equiv 0.935$                    & $0.95\pm0.07$         \\
$\psj{10}$                             & $\equiv 0.949$                    & $0.97\,^{+0.11}_{-0.10}$          & $\equiv 0.937$                    & $0.94\pm0.10$          & $\equiv 0.964$                    & $1.00\,^{+0.11}_{-0.09}$          & $\equiv 0.912$                    & $0.94\,^{+0.10}_{-0.09}$         \\
$\psj{11}$                             & $\equiv 0.926$                    & $<4.30$                           & $\equiv 0.913$                    & $0.77\,^{+0.42}_{-0.37}$          & $\equiv 0.943$                    & $2.60\,^{+2.01}_{-2.52}$          & $\equiv 0.889$                    & $2.57\,^{+1.96}_{-2.17}$         \\
$\psj{12}$                             & $\equiv 0.815$                    & nb                                & $\equiv 0.799$                    & $<3.32$                           & $\equiv 0.841$                    & nb                                & $\equiv 0.780$                    & nb                               \\
\hline
\end{tabular}}
\caption{\textbf{Effective Number of Relativistic Species} Constraints on cosmological parameters from 
Planck TT,TE,EE+lowP dataset
alone and in combination with the matter power spectrum
shape measurements from WiggleZ (MPkW),
the BAO data and the lensing constraints from Planck,
in the \lcdm + \neff\ model (\emph{nb} refers to \emph{no bound}).
For each combination,
we report the limits obtained for the two parameterizations
of the primordial power spectrum,
namely the power-law model (first column)
and the polynomial expansion (second column of each pair).
Limits are at 95\% CL around the mean value
of the posterior distribution.
For each dataset, in the case of power-law model,
the values of \psj{i} are computed according to
Eq.~\eqref{eq:psjbf}.
}\label{tab:nnu_pol}
\end{table*}

\begin{table*}[!ht]
\scalebox{0.8}{
\begin{tabular}{|c||c|c||c|c||c|c||c|c|}
\hline
Parameter	&\multicolumn{2}{c||}{Planck TT+lowP}	
&\multicolumn{2}{c||}{Planck TT+lowP+MPkW}	
&\multicolumn{2}{c||}{Planck TT+lowP+BAO} 
&\multicolumn{2}{c|}{Planck TT+lowP+lensing}\\
\hline\hline
$\Omega_b h^2$                        & $0.02214_{-0.00052}^{+0.00054}$ & $0.02127_{-0.0010}^{+0.00097}$ & $0.02217_{-0.00047}^{+0.00045}$ & $0.02161_{-0.00103}^{+0.00089}$ & $0.02228_{-0.00039}^{+0.00041}$ & $0.02219_{-0.00061}^{+0.00063}$ & $0.02210_{-0.00057}^{+0.00048}$  & $ 0.02136_{-0.00079}^{+0.00081}$\\
$\Omega_c h^2$                        & $0.1202_{-0.0047}^{+0.0044}$    & $0.1253_{-0.0064}^{+0.0067}$    & $0.1200_{-0.0039}^{+0.0043}$     & $0.1239_{-0.0067}^{+0.0072}$    & $0.1188_{-0.0029}^{+0.0028}$    & $0.1186_{-0.0033}^{+0.0032}$    & $0.1197_{-0.0042}^{+0.0041}$     & $0.1243_{-0.0059}^{+0.0060}$\\
$100\theta$                           & $1.0407_{-0.0011}^{+0.0010}$    & $1.03978_{-0.00136}^{+0.00137}$               & $1.04076_{-0.00097}^{+0.00095}$ & $1.04016_{-0.00146}^{+0.00131}$ & $1.04098_{-0.00082}^{+0.00086}$ & $1.04097_{-0.00081}^{+0.00084}$ & $ 1.04078_{-0.00091}^{+0.00096}$ & $ 1.03989_{-0.00112}^{+0.00118}$	\\
$\tau$                                & $0.080\pm0.038$                 & $0.075_{-0.044}^{+0.048}$       & $0.077_{-0.037}^{+0.038}$       & $0.077_{-0.043}^{+0.050}$       & $0.082_{-0.037}^{+0.038}$       & $0.093_{-0.047}^{+0.050}$       & $0.072_{-0.032}^{+0.034}$        & $ 0.071_{-0.037}^{+0.040}$\\
$\mnu [eV]$                     & $<0.754$                        & $<2.156$                        & $<0.455$                        & $<1.1497$                       & $<0.220$                        & $<0.261$                        & $<0.618$                         & $<1.636$\\
\hline
\end{tabular}}
\caption{\textbf{Massive Neutrinos} As Tab.~\ref{tab:nnu}, but for the \lcdm + $\mnu$ model.}\label{tab:mnuTT}
\end{table*}

\begin{table*}[!ht]
\scalebox{0.72}{\begin{tabular}{|c||c|c||c|c||c|c||c|c|}
\hline
Parameter	&\multicolumn{2}{c||}{Planck TT,TE,EE+lowP}	&\multicolumn{2}{c||}{Planck TT,TE,EE+lowP}	&\multicolumn{2}{c||}{Planck TT,TE,EE+lowP}	&\multicolumn{2}{c|}{Planck TT,TE,EE+lowP}\\
	&\multicolumn{2}{c||}{}	&\multicolumn{2}{c||}{+MPkW}	&\multicolumn{2}{c||}{+BAO}	&\multicolumn{2}{c|}{+lensing}\\
\hline\hline
$\Omega_b h^2$                        & $0.02221_{-0.00034}^{+0.00032}$   & $0.022080_{-0.00040}^{+0.00039}$ & $0.02223_{-0.00027}^{+0.00028}$ & $0.02209_{-0.00038}^{+0.00037}$ & $0.022229\pm0.00027$            & $0.02226\pm0.00033$             & $0.02215\pm0.00033$              & $ 0.02203\pm0.00041$\\
$\Omega_c h^2$                        & $0.1200_{-0.0030}^{+0.0031}$      & $0.1212_{-0.0034}^{+0.0035}$     & $0.1199_{-0.0027}^{+0.0028}$     & $0.1212_{-0.0033}^{+0.0035}$    & $0.1192\pm0.0023$               & $0.1191_{-0.0025}^{+0.0024}$    & $0.1201\pm0.0030$                & $0.1207_{-0.0035}^{+0.0033}$\\
$100\theta$                           & $1.04068_{-0.00069}^{+0.00067}$   & $1.0405_{-0.00071}^{+0.0069}$    & $1.04072_{-0.00063}^{+0.00064}$ & $1.04057_{-0.00069}^{+0.00065}$ & $1.04083_{-0.00060}^{+0.00058}$ & $1.04083_{-0.00059}^{+0.00058}$ & $ 1.04063_{-0.00069}^{+0.00068}$ & $ 1.04045_{-0.00073}^{+0.00072}$	\\
$\tau$                                & $0.081_{-0.034}^{+0.033}\pm0.038$ & $0.085_{-0.040}^{+0.042}$        & $0.080\pm0.034$                 & $0.088\pm0.037$                 & $0.083_{-0.032}^{+0.033}$       & $0.088_{-0.040}^{+0.045}$       & $0.076_{-0.032}^{+0.033}$        & $ 0.082\pm0.035$\\
$\mnu [eV]$                           & $<0.497$                          & $<0.880$                         & $<0.349$                        & $<0.458$                        & $<0.175$                        & $<0.218$                        & $<0.625$                         & $<1.172$\\
\hline
\end{tabular}}
\caption{\textbf{Massive Neutrinos} As Tab.~\ref{tab:nnu_pol}, but for the \lcdm + $\mnu$ model.}
\label{tab:mnuTTTEEE}
\end{table*}

\begin{table*}[!ht]
\scalebox{0.8}{
\begin{tabular}{|c||c| c||c|c||c|c||c|c|}
\hline
Parameter	
	&\multicolumn{2}{c||}{Planck TT+lowP}
	&\multicolumn{2}{c||}{Planck TT+lowP+MPkW}
	&\multicolumn{2}{c||}{Planck TT+lowP+BAO}
	&\multicolumn{2}{c| }{Planck TT+lowP+lensing}\\
\hline\hline
$\Omega_b h^2$                         & $0.02215\,^{+0.00080}_{-0.00084}$ & $0.02136\,^{+0.00125}_{-0.00112}$ & $0.02215\,^{+0.00070}_{-0.00071}$ & $0.02161\,^{+0.00095}_{-0.00109}$ & $0.02236\,^{+0.00049}_{-0.00050}$ & $0.02218\,^{+0.00065}_{-0.00061}$ & $0.02212\,^{+0.00082}_{-0.00079}$ & $0.02148\,^{+0.00097}_{-0.00089}$\\
$\Omega_c h^2$                         & $0.1205\,^{+0.0079}_{-0.0077}$    & $0.1302\,^{+0.0200}_{-0.0214}$    & $0.1203\,^{+0.0078}_{-0.0077}$    & $0.1251\,^{+0.0156}_{-0.0150}$    & $0.1207\,^{+0.0080}_{-0.0078}$    & $0.1326\,^{+0.0192}_{-0.0165}$    & $0.1201\,^{+0.0077}_{-0.0074}$    & $0.1301\,^{+0.0169}_{-0.0172}$   \\
$100\theta$                            & $1.0407\,^{+0.0011}_{-0.0011}$    & $1.0395\,^{+0.0021}_{-0.0020}$    & $1.0407\,^{+0.0012}_{-0.0011}$    & $1.0402\,^{+0.0018}_{-0.0017}$    & $1.0408\,^{+0.0011}_{-0.0011}$    & $1.0397\,^{+0.0015}_{-0.0016}$    & $1.0407\,^{+0.0012}_{-0.0012}$    & $1.0395\,^{+0.0018}_{-0.0017}$   \\
$\tau$                                 & $0.081\,^{+0.043}_{-0.041}$       & $0.075\,^{+0.047}_{-0.048}$       & $0.078\,^{+0.044}_{-0.041}$       & $0.076\,^{+0.049}_{-0.044}$       & $0.085\,^{+0.039}_{-0.037}$       & $0.084\,^{+0.048}_{-0.046}$       & $0.076\,^{+0.042}_{-0.038}$       & $0.074\,^{+0.043}_{-0.041}$      \\
$\neff$                                & $3.08\,^{+0.63}_{-0.60}$          & $3.44\,^{+1.51}_{-1.62}$          & $3.06\,^{+0.60}_{-0.56}$          & $3.13\,^{+1.12}_{-1.13}$          & $3.18\,^{+0.50}_{-0.49}$          & $3.94\,^{+1.28}_{-1.05}$          & $3.07\,^{+0.62}_{-0.62}$          & $3.48\,^{+1.24}_{-1.18}$         \\
\hline
\end{tabular}}
\caption{\textbf{Effective Number of Relativistic Species and Neutrino Masses} As Tab.~\ref{tab:nnu},
but for the \lcdm + \neff\ + \mnu\ model.
}\label{tab:mnu_nnu}
\end{table*}
\begin{table*}[!ht]
\scalebox{0.8}{
\begin{tabular}{|c||c| c||c|c||c|c||c|c|}
\hline
Parameter	&\multicolumn{2}{c||}{Planck TT,TE,EE+lowP}	&\multicolumn{2}{c||}{Planck TT,TE,EE+lowP}	&\multicolumn{2}{c||}{Planck TT,TE,EE+lowP}	&\multicolumn{2}{c|}{Planck TT,TE,EE+lowP}\\
	&\multicolumn{2}{c||}{}	&\multicolumn{2}{c||}{+MPkW}	&\multicolumn{2}{c||}{+BAO}	&\multicolumn{2}{c|}{+lensing}\\
\hline\hline
$\Omega_b h^2$                         & $0.02215\,^{+0.00049}_{-0.00050}$ & $0.02202\,^{+0.00055}_{-0.00057}$ & $0.02212\,^{+0.00049}_{-0.00048}$ & $0.02200\,^{+0.00050}_{-0.00051}$ & $0.02229\,^{+0.00038}_{-0.00037}$ & $0.02233\,^{+0.00042}_{-0.00041}$ & $0.02208\,^{+0.00048}_{-0.00048}$ & $0.02196\,^{+0.00053}_{-0.00054}$\\
$\Omega_c h^2$                         & $0.1191\,^{+0.0062}_{-0.0059}$    & $0.1204\,^{+0.0072}_{-0.0074}$    & $0.1190\,^{+0.0064}_{-0.0058}$    & $0.1194\,^{+0.0068}_{-0.0066}$    & $0.1192\,^{+0.0062}_{-0.0057}$    & $0.1206\,^{+0.0068}_{-0.0068}$    & $0.1184\,^{+0.0060}_{-0.0057}$    & $0.1200\,^{+0.0076}_{-0.0072}$   \\
$100\theta$                            & $1.0408\,^{+0.0009}_{-0.0009}$    & $1.0406\,^{+0.0010}_{-0.0010}$    & $1.0408\,^{+0.0009}_{-0.0009}$    & $1.0408\,^{+0.0009}_{-0.0009}$    & $1.0409\,^{+0.0008}_{-0.0008}$    & $1.0407\,^{+0.0009}_{-0.0009}$    & $1.0409\,^{+0.0009}_{-0.0009}$    & $1.0405\,^{+0.0011}_{-0.0011}$   \\
$\tau$                                 & $0.081\,^{+0.036}_{-0.036}$       & $0.085\,^{+0.043}_{-0.041}$       & $0.079\,^{+0.036}_{-0.035}$       & $0.085\,^{+0.039}_{-0.038}$       & $0.084\,^{+0.033}_{-0.034}$       & $0.090\,^{+0.042}_{-0.042}$       & $0.071\,^{+0.036}_{-0.034}$       & $0.081\,^{+0.037}_{-0.037}$      \\
$\neff$                                & $2.98\,^{+0.40}_{-0.39}$          & $2.99\,^{+0.48}_{-0.50}$          & $2.96\,^{+0.41}_{-0.38}$          & $2.92\,^{+0.45}_{-0.42}$          & $3.04\,^{+0.37}_{-0.34}$          & $3.14\,^{+0.40}_{-0.42}$          & $2.93\,^{+0.39}_{-0.38}$          & $2.99\,^{+0.52}_{-0.51}$         \\
\hline
\end{tabular}}
\caption{\textbf{Effective Number of Relativistic Species and Neutrino Masses} As Tab.~\ref{tab:nnu_pol},
but for the  \lcdm + \neff\ + \mnu\ model.
}\label{tab:mnu_nnu_pol}
\end{table*}

\begin{table*}[!ht]
\scalebox{0.82}{\begin{tabular}{|c||c|c||c|c||c|c||c|c|}
\hline
Parameter	&\multicolumn{2}{c||}{Planck TT+lowP}	
&\multicolumn{2}{c||}{Planck TT+lowP+MPkW}	
&\multicolumn{2}{c||}{Planck TT+lowP+BAO} 
&\multicolumn{2}{c|}{Planck TT+lowP+lensing}\\
\hline\hline
$\Omega_b h^2$                        & $0.02229_{-0.00058}^{+0.00064}$ & $0.02135_{-0.00095}^{+0.00103}$ & $0.02228_{-0.00052}^{+0.00057}$ & $0.02169_{-0.00098}^{+0.00088}$ & $0.02252_{-0.00047}^{+0.00051}$ & $0.02281_{-0.00062}^{+0.00066}$ & $0.02230_{-0.00063}^{+0.00062}$  & $ 0.02157_{-0.00087}^{+0.00090}$\\    
$\Omega_c h^2$                        & $0.1217_{-0.0105}^{+0.0090}$    & $0.1363_{-0.0158}^{+0.0171}$    & $0.1209_{-0.0118}^{+0.0083}$    & $0.1315_{-0.0144}^{+0.0149}$    & $0.1214\pm0.0081$               & $0.1359_{-0.0195}^{+0.0210}$    & $0.1226_{-0.0069}^{+0.0071}$     & $0.1339_{-0.0139}^{+0.0152}$\\
$100\theta$                           & $1.04039_{-0.00105}^{+0.00106}$ & $1.0399\pm0.00170$              & $1.04048_{-0.00118}^{+0.00100}$ & $1.03940_{-0.00161}^{+0.00151}$ & $1.04064_{-0.00105}^{+0.00096}$ & $1.03948_{-0.00158}^{+0.00161}$ & $ 1.04040_{-0.00107}^{+0.00098}$ & $ 1.03904_{-0.00148}^{+0.00139}$	\\
$\tau$                                & $0.088_{-0.041}^{+0.043}$       & $0.077\pm0.046$                 & $0.077_{-0.038}^{+0.039}$          & $0.077_{-0.046}^{+0.048}$       & $0.095_{-0.040}^{+0.041}$       & $0.087_{-0.045}^{+0.049}$       & $0.084\pm0.039$                  & $ 0.075\pm0.043$\\
$\mnu [eV]$                     & $<0.676$                        & $<2.449$                        & $<0.4778$                       & $<1.167$                        & $<0.263$                        & $<0.481$                        & $<0.641$                         & $<1.385$\\
\hline
\end{tabular}}
\caption{\textbf{Massive neutrinos and extra massive sterile neutrino species} As Tab.~\ref{tab:nnu},
but for the \lcdm + $\mnu$ + $\neff$ + $\meff$ model.}\label{tab:meffTT}
\end{table*}
\begin{table*}[!ht]
\scalebox{0.8}{\begin{tabular}{|c||c|c||c|c||c|c||c|c|}
\hline
Parameter	&\multicolumn{2}{c||}{Planck TT,TE,EE+lowP}	&\multicolumn{2}{c||}{Planck TT,TE,EE+lowP}	&\multicolumn{2}{c||}{Planck TT,TE,EE+lowP}	&\multicolumn{2}{c|}{Planck TT,TE,EE+lowP}\\
	&\multicolumn{2}{c||}{}	&\multicolumn{2}{c||}{+MPkW}	&\multicolumn{2}{c||}{+BAO}	&\multicolumn{2}{c|}{+lensing}\\
\hline\hline
$\Omega_b h^2$                        & $0.02286_{-0.00036}^{+0.00037}$ & $0.02226\pm0.00046$  & $0.02226_{-0.00038}^{+0.00036}$ & $0.02184_{-0.00039}^{+0.00041}$ & $0.02241_{-0.00032}^{+0.00036}$ & $0.02242_{-0.00038}^{+0.00041}$ & $ 0.02227_{-0.00036}^{+0.00037}$ & $0.02226_{-0.00047}^{+0.00049}$\\
$\Omega_c h^2$                        & $0.1207_{-0.0071}^{+0.0061}$    & $0.1223_{-0.0076}^{+0.0074}$     & $0.1189_{-0.0099}^{+0.0080}$    & $0.1203_{-0.0110}^{+0.0083}$    & $0.1189_{-0.0081}^{+0.0068}$    & $0.1201_{-0.0079}^{+0.0077}$    & $0.1207_{-0.0070}^{+0.0064}$     & $0.1228_{-0.0063}^{+0.0068}$\\
$100\theta$                           & $1.04043_{-0.00074}^{+0.00071}$ & $1.040217_{-0.00084}^{+0.00078}$ & $1.04048_{-0.00073}^{+0.00070}$ & $1.04038_{-0.00079}^{+0.00073}$ & $1.04070_{-0.00071}^{+0.00066}$ & $1.04058_{-0.00081}^{+0.00073}$ & $1.04046_{-0.00075}^{+0.00069}$  & $1.04009_{-0.00084}^{+0.00085}$	\\
$\tau$                                & $0.087_{-0.036}^{+0.035}$       & $0.090_{-0.043}^{+0.044}$        & $0.084_{-0.033}^{+0.034}$       & $0.088_{-0.039}^{+0.040}$       & $0.089\pm0.034$                 & $0.092_{-0.043}^{+0.042}$       & $0.079_{-0.031}^{+0.034}$        & $0.0083_{-0.037}^{+0.036}$\\
$\mnu [eV]$                     & $<0.528$                        & $<0.829$                         & $<0.365$                        & $<0.401$                        & $<0.199$                        & $<0.219$                        & $<0.573$                         & $<1.071$\\
\hline
\end{tabular}}
\caption{\textbf{Massive neutrinos and extra massive sterile neutrino species} As Tab.~\ref{tab:nnu_pol},
but for the \lcdm + $\mnu$ + $\neff$ + $\meff$ model.}\label{tab:meffTTTEEE}
\end{table*}

\begin{table*}[!ht]
\scalebox{0.82}{\begin{tabular}{|c||c|c||c|c||c|c||c|c|}
\hline
Parameter	&\multicolumn{2}{c||}{Planck TT+lowP}	&\multicolumn{2}{c||}{Planck TT+lowP+MPkW}	&\multicolumn{2}{c||}{Planck TT+lowP+BAO} &\multicolumn{2}{c|}{Planck TT+lowP+lensing}\\
\hline\hline
$\Omega_b h^2$	&$0.02245_{-0.00046}^{+0.00048}$	&$0.02178_{-0.00079}^{+0.00080}$	
&$0.02240_{-0.00043}^{+0.00045}$	&$0.02191_{-0.00071}^{+0.00074}$	
&$0.02248_{-0.00040}^{+0.00043}$	&$0.02224_{-0.00062}^{+0.00065}$
&$0.02245_{-0.00047}^{+0.00046}$	&$0.02182_{-0.00077}^{+0.00082}$\\

$\Omega_c h^2$	&$0.1229_{-0.0047}^{+0.0049}$	&$0.1267_{-0.0061}^{+0.0062}$	
&$0.1234_{-0.0043}^{+0.0045}$	&$0.1262_{-0.0056}^{+0.0058}$	
&$0.1219_{-0.0028}^{+0.0027}$	&$0.1222\pm0.0032$
&$0.1222_{-0.0044}^{+0.0043}$	&$0.1253_{-0.0059}^{+0.0058}$\\

$100\theta$	&$1.04045_{-0.00097}^{+0.00096}$	&$1.0399\pm0.0011$	
&$1.04045_{-0.00097}^{+0.00094}$	&$1.0401_{-0.0011}^{+0.0010}$	
&$1.04065_{-0.00080}^{+0.00081}$	&$1.04060_{-0.00085}^{+0.00084}$
&$1.04059_{-0.00092}^{+0.00091}$	&$1.0401\pm0.0010$	\\

$\tau$	&$0.088_{-0.038}^{+0.039}$		&$0.074_{-0.043}^{+0.047}$	
&$0.084_{-0.039}^{+0.040}$	&$0.076_{-0.043}^{+0.049}$	
&$0.090\pm0.038$	&$0.091_{-0.043}^{+0.046}$
&$0.078\pm0.034$		&$0.062_{-0.037}^{+0.038}$\\

$m_a [eV]$	&$<1.97$	& nb	
&$<1.09$	&$<1.63$	
&$<0.929$	&$<1.07$
&$<1.45$	& $<2.15$\\

$n_S$	&$0.974_{-0.015}^{+0.014}$	&--	
&$0.974\pm0.012$	&--	
&$0.978\pm0.010$	&--
&$0.977_{-0.013}^{+0.012}$	&--	\\

$\ln[10^{10}A_s]$	&$3.119_{-0.074}^{+0.075}$	&--	
&$3.112\pm0.077$	&--	
&$3.121_{-0.075}^{+0.076}$	&--
&$3.096_{-0.061}^{+0.062}$	&--	\\
\hline
\end{tabular}•}
\caption{\textbf{Thermal Axion} As Tab.~\ref{tab:nnu},
but for the \lcdm + $m_a$ model.}\label{tab:maTT}
\end{table*}•
\begin{table*}[!ht]
\scalebox{0.77}{\begin{tabular}{|c||c| c||c|c||c|c||c|c|}
\hline
Parameter	&\multicolumn{2}{c||}{Planck TT,TE,EE+lowP}	&\multicolumn{2}{c||}{Planck TT,TE,EE+lowP}	&\multicolumn{2}{c||}{Planck TT,TE,EE+lowP}	&\multicolumn{2}{c|}{Planck TT,TE,EE+lowP}\\
	&\multicolumn{2}{c||}{}	&\multicolumn{2}{c||}{+MPkW}	&\multicolumn{2}{c||}{+BAO}	&\multicolumn{2}{c|}{+lensing}\\
\hline\hline
$\Omega_b h^2$	&$0.02248\pm0.00032$	&$0.02241_{-0.00038}^{+0.00039}$	
&$0.02245_{-0.00031}^{+0.00030}$	&$0.02236_{-0.00038}^{+0.00037}$	
&$0.02250_{-0.00029}^{+0.00030}$	&$0.02248_{-0.00036}^{+0.00038}$
&$0.02248_{-0.00030}^{+0.00033}$	&$0.02242_{-0.00038}^{+0.00039}$	\\

$\Omega_c h^2$	&$0.1232_{-0.0036}^{+0.0034}$	&$0.1233_{-0.0043}^{+0.0041}$	
&$0.1236_{-0.0033}^{+0.0032}$	&$0.1241_{-0.0040}^{+0.0037}$	
&$0.1224_{-0.0024}^{+0.0023}$	&$0.1223\pm0.0029$
&$0.1231_{-0.0033}^{+0.0032}$	&$0.1224_{-0.0043}^{+0.0039}$	\\

$100\theta$	&$1.04030_{-0.00067}^{+0.00065}$	&$1.04020_{-0.00064}^{+0.00066}$	
&$1.04029\pm0.00065$	&$1.04022_{-0.00067}^{+0.00068}$	
&$1.04047_{-0.00060}^{+0.00059}$	&$1.04045_{-0.00057}^{+0.00059}$
&$1.04037\pm0.00066$	&$1.04027\pm0.00063$	\\

$\tau$	&$0.090_{-0.034}^{+0.033}$		&$0.090_{-0.042}^{+0.043}$	
&$0.087\pm0.034$	&$0.091\pm 0.039$	
&$0.092\pm0.034$	&$0.093_{-0.042}^{+0.043}$
&$0.075\pm0.028$		&$0.071_{-0.032}^{+0.034}$	\\

$m_a [eV]$	&$<2.09$	& $<2.44$	
&$<1.19$	&$<1.90$	
&$<0.741$	&$<1.19$
&$<1.68$	& $<2.44$\\

$n_S$	&$0.972_{-0.012}^{+0.011}$	&--	
&$0.9734\pm0.0098$	&--	
&$0.9754_{-0.0089}^{+0.0092}$	&--
&$0.974_{-0.011}^{+0.010}$	&--	\\
\hline
\end{tabular}•}
\caption{\textbf{Thermal Axion} As Tab.~\ref{tab:nnu_pol},
but for the \lcdm + $m_a$ model.}\label{tab:maTTTEEE}
\end{table*}

\end{document}